\documentclass[12pt,preprint]{aastex}

 \usepackage{epstopdf}

 \slugcomment{Accepted Version 1}

\newcommand\pv{\mbox{$p_{V}$}}
\newcommand\irfactor{\mbox{$p_{IR}/p_{V}$}}

\begin{document}

 \DeclareGraphicsExtensions{.pdf,.gif,.jpg}

 \title{NEOWISE Observations of Near-Earth Objects: Preliminary Results}
\author{A. Mainzer\altaffilmark{1}$^{,}$\altaffilmark{27}$^{,}$\altaffilmark{28}$^{,}$\altaffilmark{29}, T. Grav\altaffilmark{2}$^{,}$\altaffilmark{27}$^{,}$\altaffilmark{28}, J. Bauer\altaffilmark{1}$^{,}$\altaffilmark{3}$^{,}$\altaffilmark{27}$^{,}$\altaffilmark{28}, J. Masiero\altaffilmark{1}$^{,}$\altaffilmark{27}$^{,}$\altaffilmark{28}$^{,}$\altaffilmark{29}, R. S. McMillan\altaffilmark{4}$^{,}$\altaffilmark{27}$^{,}$\altaffilmark{28}$^{,}$\altaffilmark{29}, R. M. Cutri\altaffilmark{3}, R. Walker\altaffilmark{5}, E. Wright\altaffilmark{6}, P. Eisenhardt\altaffilmark{1}, D. J. Tholen\altaffilmark{7}, T. Spahr\altaffilmark{8}, R. Jedicke\altaffilmark{7}, L. Denneau\altaffilmark{7}, E. DeBaun\altaffilmark{9}, D. Elsbury\altaffilmark{10}, T. Gautier\altaffilmark{11}, S. Gomillion\altaffilmark{12}, E. Hand\altaffilmark{13}, W. Mo\altaffilmark{2}, J. Watkins\altaffilmark{14}, A. Wilkins\altaffilmark{15}, G. L. Bryngelson\altaffilmark{16}$^{,}$\altaffilmark{27}$^{,}$\altaffilmark{29}, A. Del Pino Molina\altaffilmark{17}$^{,}$\altaffilmark{28}, S. Desai\altaffilmark{18}$^{,}$\altaffilmark{27}, M. G\'{o}mez Camus\altaffilmark{19}$^{,}$\altaffilmark{28}, S. L. Hidalgo\altaffilmark{17}$^{,}$\altaffilmark{28}, I. Konstantopoulos\altaffilmark{20}$^{,}$\altaffilmark{27}$^{,}$\altaffilmark{29}, J. A. Larsen\altaffilmark{21}$^{,}$\altaffilmark{27}$^{,}$\altaffilmark{28}$^{,}$\altaffilmark{29}, C. Maleszewski\altaffilmark{4}$^{,}$\altaffilmark{27}$^{,}$\altaffilmark{28}$^{,}$\altaffilmark{29}, M. A. Malkan\altaffilmark{6}$^{,}$\altaffilmark{27}$^{,}$\altaffilmark{29}, J.-C. Mauduit\altaffilmark{22}$^{,}$\altaffilmark{27}$^{,}$\altaffilmark{29}, B. L. Mullan\altaffilmark{20}$^{,}$\altaffilmark{28}, E. W. Olszewski\altaffilmark{23}$^{,}$\altaffilmark{28}, J. Pforr\altaffilmark{24}$^{,}$\altaffilmark{28}, A. Saro\altaffilmark{25}$^{,}$\altaffilmark{28}, J. V. Scotti\altaffilmark{4}$^{,}$\altaffilmark{27}$^{,}$\altaffilmark{28}$^{,}$\altaffilmark{29}, L. H. Wasserman\altaffilmark{26}$^{,}$\altaffilmark{28}}

 \altaffiltext{1}{Jet Propulsion Laboratory, California Institute of Technology, Pasadena, CA 91109 USA}
 \altaffiltext{2}{Johns Hopkins University, Baltimore, MD}
\altaffiltext{3}{Infrared Processing and Analysis Center, California Institute of Technology, Pasadena, CA 91125, USA}
\altaffiltext{4}{Lunar and Planetary Laboratory, University of Arizona, 1629 East University Blvd., Tucson, AZ 85721-0092, USA}
\altaffiltext{5}{Monterey Institute for Research in Astronomy, Monterey, CA USA}
\altaffiltext{6}{Department of Physics and Astronomy, UCLA, PO Box 91547, Los Angeles, CA 90095-1547 USA}
\altaffiltext{7}{Institute for Astronomy, University of Hawaii, 2680 Woodlawn Drive, Honolulu, HI USA}
\altaffiltext{8}{Minor Planet Center, Harvard-Smithsonian Center for Astrophysics, 60 Garden Street, Cambridge, MA 02138 USA}
\altaffiltext{9}{Dartmouth University, Hanover, NH 03755 USA}
\altaffiltext{10}{University of California Santa Barbara, Broida Hall, Santa Barbara, CA 93103 USA }
\altaffiltext{11}{Cornell University, Ithaca, NY 14853 USA}
\altaffiltext{12}{Embry-Riddle Aeronautical University, 600 S. Clyde Morris Boulevard, Daytona Beach, FL 32114 USA}
\altaffiltext{13}{University of Missouri-Kansas City, Kansas City, MO 64110 USA}
\altaffiltext{14}{Department of Earth and Space Sciences, UCLA, 595 Charles Young Drive East, Box 951567, Los Angeles, CA 90095 USA}
\altaffiltext{15}{Department of Astronomy, University of Maryland, College Park, MD 20742 USA}
\altaffiltext{16}{Department of Physics and Astronomy, Clemson University, 8304 University Station, Clemson, SC 29634 USA}
\altaffiltext{17}{Instituto de Astrof\`{i}sica de Canarias, V\`{i}a Lactea, E38200-La Laguna, Tenerife, Canary Islands, Spain}
\altaffiltext{18}{University of Illinois, Urbana-Champagne, National Center for Supercomputing Applications, 1205 W. Clark St., Urbana, IL 61801, USA}
\altaffiltext{19}{Departmento de Ciencias Fisicas, Facultad de Ingenieria, Universidad Andres Bello, Republica 220, Santiago, Chile}
\altaffiltext{20}{Department of Astronomy and Astrophysics, Penn State University, 525 Davey Lab, University Park, PA 16802 USA}
\altaffiltext{21}{Department of Physics, United States Naval Academy, Annapolis, MD 21402 USA}
\altaffiltext{22}{Infrared Processing and Analysis Center/Spitzer Science Center, California Institute of Technology, Mail Code 220-6, Pasadena, CA 91125 USA}
\altaffiltext{23}{Steward Observatory, University of Arizona, Tucson, AZ 85721 USA}
\altaffiltext{24}{Institute of Cosmology and Gravitation, University of Portsmouth, Dennis Sciama Building, Burnaby Road, Portsmouth PO1 3FX, UK}
\altaffiltext{25}{Department of Physics, Ludwig-Maximilians-Universitat, Scheinerstr. 1, 81679, Munich, Germany}
\altaffiltext{26}{Lowell Observatory, 1400 W. Mars Hill Road, Flagstaff, AZ 86001 USA}
\altaffiltext{27}{Visiting Astronomer, Kitt Peak National Observatory, National Optical Astronomy Observatory, which is operated by the Association of Universities for Research in Astronomy (AURA) under cooperative agreement with the National Science Foundation.}
\altaffiltext{28}{Visiting astronomer, Cerro Tololo Inter-American Observatory, National Optical Astronomy Observatory, which are operated by the Association of Universities for Research in Astronomy, under contract with the National Science Foundation.}
\altaffiltext{29}{The WIYN Observatory is a joint facility of the University of Wisconsin-Madison, Indiana University, Yale University, and the National Optical Astronomy Observatory.}

 \email{amainzer@jpl.nasa.gov}

 \begin{abstract}
 With the NEOWISE portion of the \emph{Wide-field Infrared Survey Explorer} (WISE) project, we have carried out a highly uniform survey of the near-Earth object (NEO) population at thermal infrared wavelengths ranging from 3 to 22 $\mu$m, allowing us to refine estimates of their numbers, sizes, and albedos.  The NEOWISE survey detected NEOs the same way whether they were previously known or not, subject to the availability of ground-based follow-up observations, resulting in the discovery of more than 130 new NEOs.  The survey's uniform sensitivity, observing cadence, and image quality have permitted extrapolation of the 428 near-Earth asteroids (NEAs) detected by NEOWISE during the fully cryogenic portion of the WISE mission to the larger population. We find that there are 981$\pm$19 NEAs larger than 1 km and 20,500$\pm$3000 NEAs larger than 100 m.  We show that the Spaceguard goal of detecting 90\% of all 1 km NEAs has been met, and that the cumulative size distribution is best represented by a broken power law with a slope of 1.32$\pm$0.14 below 1.5 km.  This power law slope produces $\sim13,200\pm$1,900 NEAs with $D>$140 m. Although previous studies predict another break in the cumulative size distribution below $D\sim$50-100 m, resulting in an increase in the number of NEOs in this size range and smaller, we did not detect enough objects to comment on this increase. The overall number for the NEA population between 100-1000 m is lower than previous estimates.  The numbers of near-Earth comets and potentially hazardous NEOs will be the subject of future work.  
 \end{abstract}

 \section{Introduction}
The near-Earth objects (NEOs) are a population of interest for a wide range of scientific investigations and practical considerations.  The NEO population (defined as asteroids or comets with perihelion distances $q\leq1.3$ AU; \emph{http://neo.jpl.nasa.gov/neo/groups.html}) is thought to be made up of both asteroids and comets ranging in size from objects tens of kilometers in diameter \citep{Shoemaker} down to dust grains.  Although nearly $\sim$8,000 NEOs have been discovered to date at all size ranges (\emph{http://neo.jpl.nasa.gov/stats}), this number represents only a fraction of the total population thought to exist \citep{Bottke}.  Insights into several fundamental questions in planetary science can be gained from large-scale studies of NEO orbital distributions and physical properties such as diameters and albedos.  

\subsection{Origins and Evolution of NEOs}
NEOs are a transitory population, with dynamic lifetimes of $10^6$ to $10^8$ years \citep{MorbidelliGladman}.  They must therefore be continually replenished in order to maintain the population we observe today.  It has been shown that NEOs are likely to have been delivered primarily from specific regions \citep{Wetherill88, Rabinowitz97a, Rabinowitz97b, Bottke} such as the $\nu_6$ secular resonance and 3:1 mean motion resonance with Jupiter.  The near-Earth asteroids (NEAs) in the NEO population are thought to have been pushed by the Yarkovsky thermal drift force from different parts of the Main Belt into resonances that may preferentially inject them into the inner Solar System \citep{Bottke}.  Near-Earth comets (NECs) also constitute a part of the NEO population and are thought to originate from a variety of regions, including the Kuiper belt, the scattered disk, the Oort cloud, and the Jupiter Trojan clouds \citep{LevisonDuncan, DuncanLevison, Weissman96, Levison97}.  Efforts to more precisely constrain the origins of the NEOs have been hampered by the relative paucity of observational data.  Surveys such as the Lincoln Near-Earth Asteroid Research Program, the Near Earth Asteroid Tracking Program, the Lowell Near-Earth Object Survey, the Catalina Sky Survey, PanSTARRS, and Spacewatch \citep{Stokes,Helin,Koehn,Larson,McMillan} have identified NEOs, but they observe in visible wavelengths, so they are not able to sense low albedo NEOs as effectively, nor can visible light measurements provide strong constraints on diameters, albedos, or other physical characteristics. Size and albedo distributions for NEOs derived from these visible light surveys remain uncertain (particularly for sub-kilometer objects) since these models rely upon the absolute magnitude ($H$) through an assumed albedo to derive size \citep{StuartBinzel, Morbi02, Shoemaker90, LuuJewitt}.  Studies linking meteorite falls to NEO source regions in the main asteroid belt have been performed \citep[e.g.][]{ThomasBinzel, Vernazza}, but meteorites have been heavily processed by passage through the Earth's atmosphere, which also introduces biases in composition and size.  

 \citet{Bottke} were limited in their ability to model the NEO population by the fact that the Spacewatch data upon which their results were based provided them only a very small sample of optically-selected NEOs.  The paucity of physical characterization data for large numbers of NEOs has complicated efforts to identify weaker source regions. Models of NEO origins such as \citet{Bottke} can be tested and improved by obtaining a sample of NEOs with accurately measured diameters and albedos, well-known orbits, and well-understood survey detection biases and efficiency. Infrared observations of all classes of minor planets are useful for determining size and albedo distributions, as well as thermophysical properties such as thermal inertia, the magnitude of non-gravitational forces, and surface roughness \citep[e.g.,][]{Bhattacharya,Tedesco02,Trilling,Harris09}. 

\subsection{NEOs As Impactors}
Asteroids and comets have impacted Earth throughout its history, profoundly altering the course of life on Earth \citep{Alvarez,Hildebrand}.  Currently available estimates of the size distribution of potentially hazardous NEOs lead to the assessment that objects capable of causing global catastrophe (larger than 1 km in diameter) are thought to impact approximately every 700,000 years \citep{NRC}.  However, smaller objects can still cause considerable damage \citep{NEOSDT,Chapman}.  Recent simulations suggest that the 1908 Tunguska impact could have been caused by an object as small as 30-50 m \citep{NRC}, which would imply that the potential for damage caused by smaller objects has been underestimated.  However, this increased risk from smaller impacts is offset by recent population estimates \citep{BosloughHarris,Harris08} that indicate that the population of asteroids in the size range between several tens to hundreds of meters in diameter may be as much as a factor of three less than estimated using a straight-line power law of slope -2.354 as proposed by \citet{Bottke}, or a slope of -1.87. Nevertheless, considerable uncertainty about the true NEO size distribution persists due to the lack of accurate diameters, albedos, and surveys that are not substantially biased against low albedo objects.  Even less is known about the subset of NEOs that are considered potentially hazardous.

\section{Observations}
WISE is a NASA Medium-class Explorer mission designed to survey the entire sky in four infrared wavelengths, 3.4, 4.6, 12 and 22 $\mu$m (denoted $W1$, $W2$, $W3$, and $W4$ respectively).  In a Sun-synchronous orbit, WISE observed the entire sky in all four filters simultaneously using three beamsplitters and a scan mirror.  The pre-launch description of the mission, including its design and construction, is given in \citet{Liu, Mainzer}, and the post-launch performance is discussed in \citet{Wright}.  The final mission data products are a multi-epoch image atlas and source catalogs that will serve as an important legacy for future research.   The survey has yielded observations of over 157,000 minor planets, including Near-Earth Objects (NEOs), Main Belt Asteroids (MBAs), comets, Hildas, Trojans, Centaurs, and scattered disk objects \citep[][hereafter M11A]{Mainzer11a}.  WISE has observed nearly two orders of magnitude more minor planets than its predecessor, the \emph{Infrared Astronomical Satellite} \citep[IRAS;][]{Tedesco, Matson}.  WISE launched on 14 December 2009.  The WISE survey began on 14 January, 2010, and the mission exhausted its primary tank cryogen on 5 August, 2010.  An augmentation to the WISE baseline data processing pipeline, ``NEOWISE," permitted a search for new moving objects to be carried out using the WISE data in near-real time.  NEOWISE also allowed the individual exposures for each part of the sky (an average of eight on the ecliptic plane, but rising into the hundreds near the ecliptic poles) to be archived and made available to the public.  Exhaustion of the secondary cryogen tank occurred on 1 October, 2010, and the survey was continued as the NEOWISE Post-Cryogenic Mission using only bands $W1$ and $W2$ until 1 February, 2011. 

\begin{figure}
\figurenum{1}
\includegraphics[width=6in]{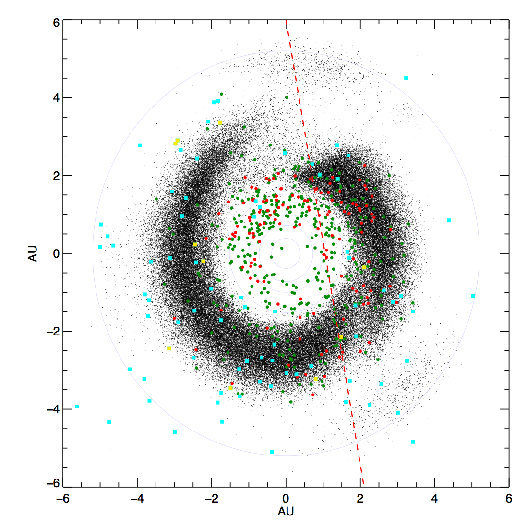}
\caption{\label{fig:xyplot}WISE always surveys near 90$^\circ$ solar elongation; this figure shows a top-down view of the objects detected by NEOWISE as of February 2011 (distances are given in AU). The outermost circle represents Jupiter's orbit; the interior circles represent the terrestrial planets.  Previously known NEOs are shown as green circles; new NEOs discovered by NEOWISE are shown as red circles; previously known comets observed by WISE are shown as cyan squares, and comets discovered by NEOWISE are shown as yellow squares.   All other objects are shown as black points. The drop in density of objects observed near (+2, +2) AU in the figure is due to the exhaustion of the secondary tank's cryogen on 5 August, 2010, resulting in the loss of band $W4$.  The dashed red line indicates the survey scan plane at the time of the exhaustion of the primary tank and the start of the NEOWISE Post-Cryogenic Mission on 1 October 2010; the survey was completed on 1 February 2011.  The drop in detections near (+2, -2) AU in the figure is due to the intersection of the galactic plane with the ecliptic plane; the higher backgrounds and confusion caused by galactic cirrus resulted in the identification of fewer sources.}
\end{figure}

As described in \citet{Wright} and M11A, the NEOWISE survey cadence resulted in most minor planets in the WISE sample receiving an average of 10-12 observations over $\sim$36 hours, although some NEOs were observed dozens or even hundreds of times. NEOWISE detected minor planets using the WISE Moving Object Processing System (WMOPS). WMOPS was adapted from the Pan-STARRS Moving Object Processing System \citep{Kubica} to account for the different cadence and observing strategy used by WISE.  During the entire period of survey operations, NEOWISE is known to have observed at least 584 NEOs and is responsible for the discovery of 135 new, previously unknown NEOs.  For purposes of determining the debiased population characteristics of the NEOs, in this paper we consider only those objects detected during the fully cryogenic portion of the survey by the WMOPS pipeline using the first-pass WISE data processing pipeline as described below.  This sample consists of a set of 428 NEOs, of which 314 were recoveries of objects discovered by other groups, and 114 were NEOWISE discoveries.  By considering this subset of NEOs, we can study a population that was observed in a more-or-less uniform fashion.  This approach facilitates the understanding of the NEOWISE survey biases, allowing us to model the unobserved portion of the NEO population.  We will consider the NEOs detected during the post-cryogenic portion of the survey in a future work.  Figure \ref{fig:xyplot} shows the objects detected by NEOWISE using the first-pass data processing pipeline; while most of the solar system was surveyed during the fully cryogenic portion of the mission, some of it was observed with only bands $W1$ and $W2$, leading to a significant drop in sensitivity.  For this and other reasons, we must debias the survey in order to understand the properties of the NEO population and disentangle them from survey biases.

The observations of the objects listed in Table 1 were retrieved by querying the Minor Planet Center's (MPC) observation files to look for all instances of individual WISE detections of the desired objects that were reported using the WISE Moving Object Processing System (WMOPS).  The resulting set of position/time pairs were used as the basis of a query of WISE source detections processed with the First Pass version of the WISE data process pipeline (Version 3.5) in individual exposures (also known as ``Level 1b" images) using the Infrared Science Archive \citep[IRSA;][]{Cutri}.  In order to ensure that only observations of the desired moving object were returned from the query, the search radius was restricted to 0.3 arcsec from the position listed in the MPC observation file.  A radius of 0.3 arcsec was chosen to account for any round-off error between the decimal format used by the WISE catalog and the hexigesimal format used by the MPC.  Additionally, since WISE collected a single exposure every 11 seconds, the modified Julian date was required to be within 2 seconds of the time specified by the MPC.  The artifact identification flag cc\_flags was allowed to be equal to either 0, P or p; this flag indicates that the sources were unlikely to have been affected by persistence. We required that the flag ph\_qual be equal to A, B, or C (this flag indicates that the source is likely to have been a valid detection).  As described in \citet[][hereafter M11B]{Mainzer11b} and \citet{Cutri}, we included observations with magnitudes close to experimentally-derived saturation limits, but when sources became brighter than $W1=6$, $W2=6$, $W3=4$ and $W4=0$, we increased the error bars on these points to 0.2 magnitudes and applied a linear correction to $W3$ (see the WISE Explanatory Supplement for details).  Each object had to be observed a minimum of three times in at least one WISE band, and to avoid having low-level noise detections and/or cosmic rays contaminating our thermal model fits, we required that observations in more than one band appear at least 40\% of the number of observations found in the band with the largest number of observations (usually $W2$ or $W3$ for NEOs). The WMOPS system is designed to reject inertially fixed objects such as stars and galaxies in bands $W3$ and $W4$. However, the source density in bands $W1$ and $W2$ is $\sim$100 times higher than in bands $W3$ and $W4$, so it was more likely that asteroid detections were confused with stars or galaxies at these wavelengths.  In order to remove such confused asteroid detections, we cross-correlated the individual Level 1b detections with the WISE atlas and daily coadd catalogs.  Objects within 6.5 arcsec (equivalent to the WISE beam size at bands $W1$, $W2$ and $W3$) of the asteroid position which appeared in the coadded source lists at least twice and which appeared more than 30\% of the total number of coverages of a given area of sky were considered to be inertially fixed sources; these asteroid detections were considered contaminated and were not used for thermal fitting.   

\begin{figure}
\figurenum{2}
\includegraphics[width=6in]{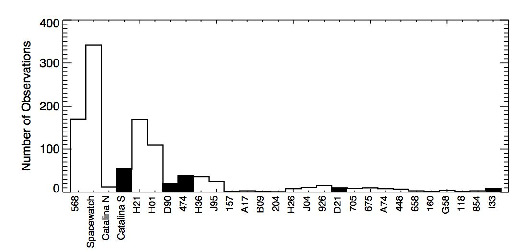}
\caption{\label{fig:obscodes}Follow-up observations within $\sim$2 weeks of the last observations of NEOs candidates detected by NEOWISE were essential for securing an orbit, as the average NEOWISE arc spanned $\sim$36 hours.  After 2 weeks, most objects detected with such arcs have sufficiently high astrometric error that they cannot be recovered.  This figure shows the number of follow-up observations contributed within 15 days of a newly discovered NEO's last observation by NEOWISE as a function of MPC observatory code.  Spacewatch (codes 291, 691, 695 and 807), H21 (the Killer Asteroid Project; \emph{http://killerasteroidproject.org/student\_obs.htm}), 568 (Mauna Kea, observers D. Tholen and students) and H01 \citep[Magdalena Ridge Observatory; ][]{Ryan} were the most prolific contributors within the first 15 days; other observers contributed follow-up over longer time spans.  Observatories in the southern hemisphere are shaded black, illustrating the paucity of southern facilities that were available for follow-up of NEOWISE discoveries. In spite of all of the follow-up observations shown in this figure, 15 NEOWISE-discovered NEOs were designated without any visible follow-up, and another $\sim$15-20 objects appeared on the MPC Confirmation Page as probable NEOs but received neither designations nor follow-up; these objects were lost.}
\end{figure}

During the cryogenic portion of the survey, 231 objects were placed on the MPC's NEO Confirmation Page (\emph{http://www.minorplanetcenter.org/iau/NEO/ToConfirm.html}).  Of these, 115 were designated as NEOs, while the remaining candidates that received visible light follow-up were confirmed to be Main Belt asteroids, Trojans, etc. (of the $\sim$300,000 tracklets submitted to the MPC, only a handful were rejected as spurious). Follow-up observations of NEOWISE-discovered NEOs were carried out by a world-wide network of amateur and professional astronomers.  Figure \ref{fig:obscodes} lists the numbers of observations of NEOWISE discoveries contributed by MPC observatory code within 15 days of the first observation by NEOWISE.  Follow-up observations were critically important for securing good long-term orbits of NEOWISE-discovered NEOs because in general, the arc lengths spanned $\sim$36 hours.  Of the 115 NEOs discovered during the cryogenic mission, 12 received designations despite having no optical follow-up. An additional 22 objects objects were deemed lost because they received no optical follow-up, and their arcs were too short to designate.  The loss of these objects will contribute some uncertainty to the size and albedo distributions computed for NEOs from NEOWISE data; this will be discussed further below.

\begin{figure}
\figurenum{3}
\includegraphics[width=6in]{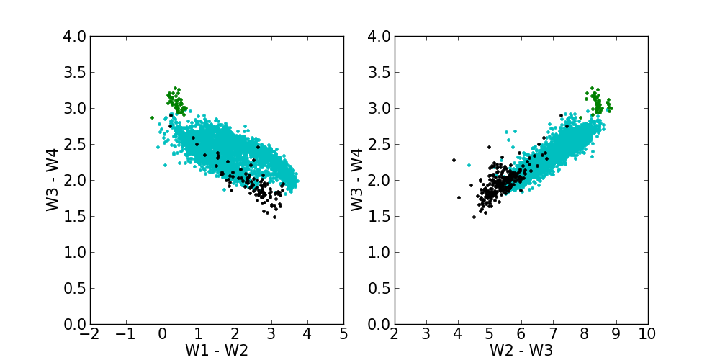}
\caption{\label{fig:colors} Left: WISE colors for objects observed in bands $W1$, $W2$, $W3$ and $W4$.  Black dots are NEOs, cyan dots are Main Belt asteroids, and green dots are Trojan asteroids. Right: WISE colors for NEOs observed in bands $W1$, $W2$, $W3$ and $W4$.}
\end{figure}

\begin{figure}
\figurenum{4}
\includegraphics[width=6in]{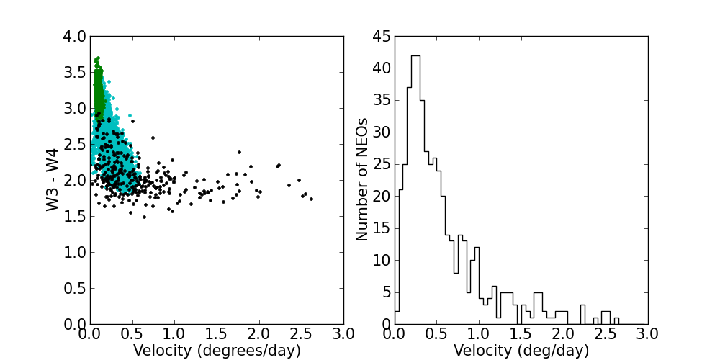}
\caption{\label{fig:Velocities} Left: $W3-W4$ color vs. average sky plane velocity for NEOs detected by NEOWISE. Right: Histogram of velocities for NEOs detected by NEOWISE. Black dots are NEOs, cyan dots are Main Belt asteroids, and green dots are Trojan asteroids. }
\end{figure}

\begin{figure}
\figurenum{5}
\includegraphics[width=6in]{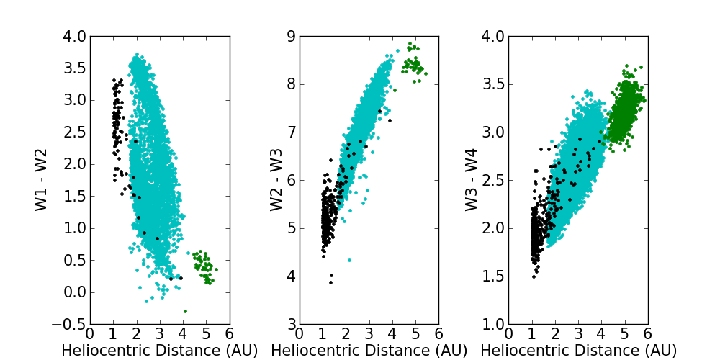}
\caption{\label{fig:distance} WISE colors vs. heliocentric distance. Black dots are NEOs, cyan dots are Main Belt asteroids, and green dots are Trojan asteroids. }
\end{figure}

Figures \ref{fig:colors} shows the typical WISE colors for NEOs, Main Belt asteroids and Trojan asteroids observed during the cryogenic NEOWISE mission.  The NEOs occupy a distinct region of color space and can readily be distinguished from other sources such as stars, distant galaxies and cool brown dwarfs.  Examination of the WISE colors can be used as a means to constrain orbits in cases where an object's orbit is uncertain (e.g. because it has only a short observational arc).  \citet{Bhattacharya} suggest that thermal colors can be used to identify new candidate minor planets; however, the vast size of the WISE single-exposure source lists (which include transient artifacts such as cosmic rays and latent images) makes techniques that focus on repeated detections of moving sources the preferred method for discovery of new objects.  Figure \ref{fig:Velocities} shows $W3-W4$ vs. apparent sky-plane velocity and the histogram of sky-plane velocities for observed NEOs; most NEOs detected by NEOWISE are moving at $\sim 0.5^\circ$/day.  Figure \ref{fig:distance} shows the WISE colors compared with heliocentric distance; as expected, the colors redden as the objects are located at greater distance from the Sun.  It may be possible to use the WISE colors and apparent sky-plane velocities to constrain orbits in cases where observational arcs are short and orbits are poorly known.  A future work will explore this further.

\section{Preliminary Thermal Modeling of NEOs}
We have created preliminary thermal models for each NEO observed by WMOPS during the fully cryogenic portion of the survey with the First Pass version of the WISE data processing pipeline described above; these thermal models will be recomputed when the final data processing is completed.  As described in M11B, we employ the spherical near-Earth asteroid thermal model (NEATM) \citep{Harris}.  The NEATM model uses the so-called beaming parameter $\eta$ to account for cases intermediate between zero thermal inertia, the Standard Thermal Model (STM) of \citet{Lebofsky_Spencer} and high thermal inertia, the Fast Rotating Model \citep[FRM;][]{Lebofsky78,Veeder89,Lebofsky_Spencer}. In the STM, $\eta$ is set to 0.756 to match the occultation diameters of (1) Ceres and (2) Pallas. In the FRM, $\eta$ is equal to $\pi$, and the temperature distribution is different than that used in NEATM, being independent of longitude.  With NEATM, $\eta$ is a free parameter that can be fit when two or more infrared bands are available (or with only one infrared band if diameter or albedo are known \emph{a priori} as is the case for objects that have been imaged by visiting spacecraft or observed with radar). 

\begin{figure}
\figurenum{6}
\includegraphics[width=3in]{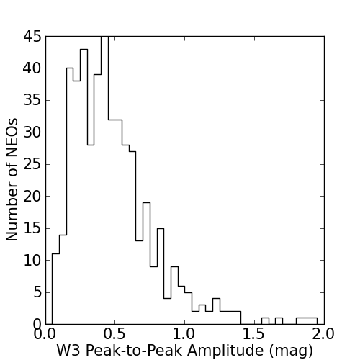}
\caption{\label{fig:lightcurves}Most NEOs detected by NEOWISE during the cryogenic portion of the mission have $W3$ peak-to-peak amplitudes equal to $\sim$0.4 magnitudes; however, $\sim$18 NEOs have $W3$ peak-to-peak amplitudes $>1$.  These objects are likely to be highly elongated and possibly binary.}
\end{figure}

Each object was modeled as a set of triangular facets covering a spherical surface with a variable diameter \citep[c.f.][]{Kaasalainen}. Although many (if not most) NEOs are non-spherical, the WISE observations generally consisted of $\sim$10-12 observations per object uniformly distributed over $\sim$36 hours (M11A), so on average, a wide range of rotational phases were sampled.  Even though this sampling helps to average out the effects of a rotating non-spherical object, caution must be exercised when interpreting the meaning of an effective diameter in these cases.   Figure \ref{fig:lightcurves} shows the average peak-to-peak amplitude in $W3$ for NEOs detected by NEOWISE during the fully cryogenic portion of the mission.  While most objects' amplitudes are $\sim$0.5 mag or less, there is a fraction of the population with extremely high amplitudes.  These objects may be far from spherical, and so the application of a spherical thermal model to these objects may not yield accurate results.  These objects are discussed in more detail below.    

 Thermal models were computed for each WISE measurement, ensuring that the correct Sun-observer-object distances were used. The temperature for each facet was computed, and the \citet{Wright} color corrections were applied to each facet. In addition, we adjusted the $W3$ effective wavelength blueward by 4\% from 11.5608$\mu$m to 11.0984 $\mu$m, the $W4$ effective wavelength redward by 2.5\% from 22.0883 $\mu$m to 22.6405 $\mu$m, and we included the -8\% and +4\% offsets to the $W3$ and $W4$ magnitude zeropoints (respectively) due to the red-blue calibrator discrepancy reported by \citet{Wright}.   The emitted thermal flux for each facet was calculated using NEATM with the bandcenters and zero points given in \citet{Wright}; the temperature at the anti-subsolar point was set to 3 K, so the facets closest to this point contribute little flux.   For NEOs, bands $W1$ and $W2$ typically contain a mix of reflected sunlight and thermal emission. The flux from reflected sunlight was computed for each WISE band as described in M11B using the IAU phase curve correction \citep{Bowell}.  Facets which were illuminated by reflected sunlight and visible to WISE were corrected with the \citet{Wright} color corrections appropriate for a G2V star.  In order to compute the fraction of the total luminosity due to reflected sunlight, it was necessary to determine the albedo in bands $W1$ and $W2$, $p_{IR}$.  This is discussed in greater detail below.

 Physical properties were computed for each object by grouping together observations with no more than a ten day gap between them.  This restriction was imposed to ensure that NEOs, which can have significant changes in distance over short times, were modeled accurately.  
  
In general, NEO absolute magnitudes ($H$) were taken from the MPC's orbital element files, and errors on $H$ were taken to be 0.3 magnitudes. However, updated $H$ magnitudes were taken from the Lightcurve Database of \citet{Warner} for 78 NEOs that were observed by WISE.  Emissivity, $\epsilon$, was assumed to be 0.9 for all wavelengths \citep[c.f.][]{Harris09}, and $G$ (the slope parameter of the magnitude-phase relationship) was set to 0.15$\pm$0.1 unless a direct measurement from \citet{Warner} was available.  We note that for some objects with high albedos, the choice of $G=0.15$ may not be appropriate; this is discussed in more detail below.
 
\begin{figure}
\figurenum{7}
\includegraphics[width=6in]{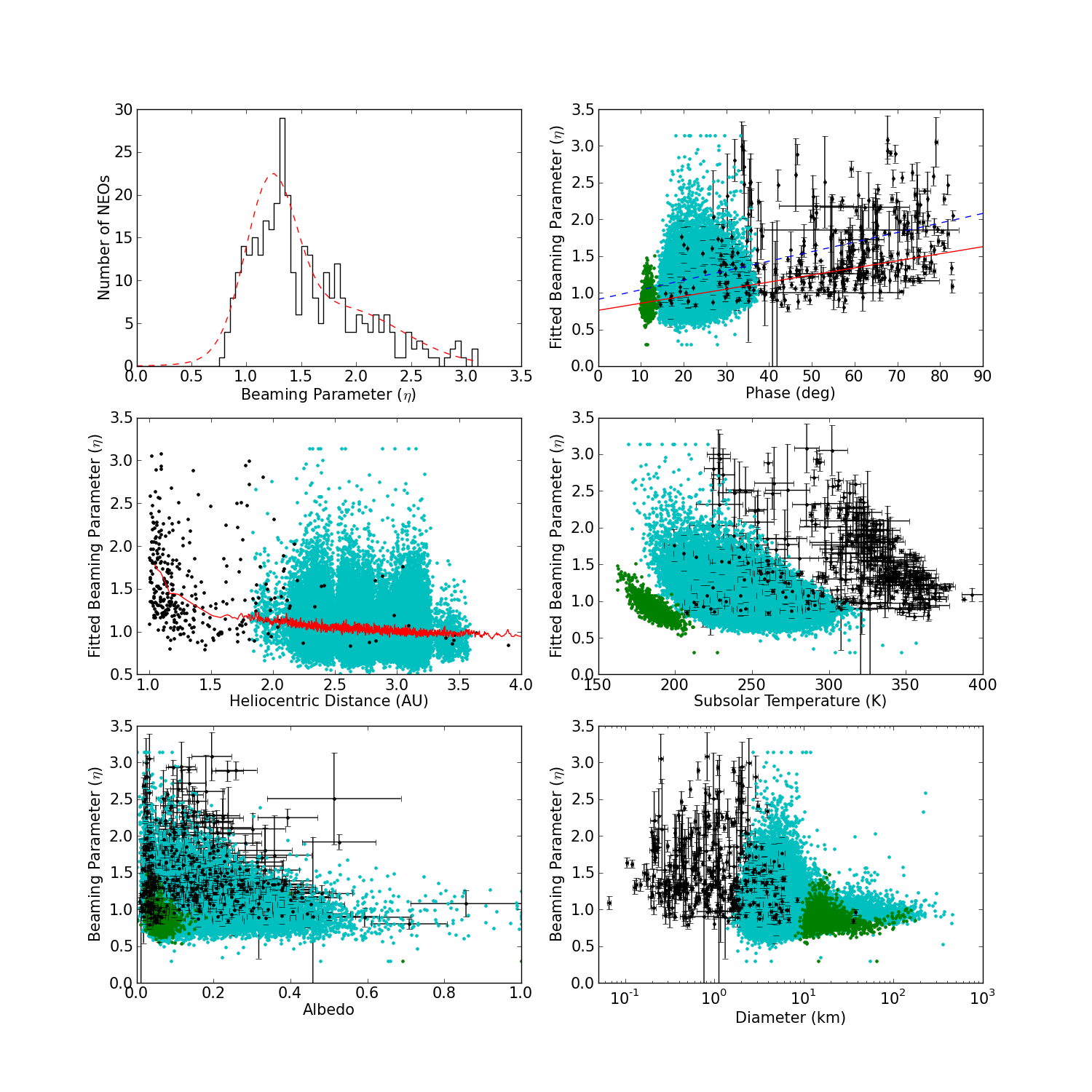}
\caption{\label{fig:beaming}Top Left: Histogram of beaming parameter ($\eta$) values for the NEOs detected by the NEOWISE WMOPS pipeline during the first-pass processing carried out throughout the cryogenic portion of the mission. The mean value of $\eta$ for objects with fitted $\eta$ was 1.4$\pm$0.5. Top Right: The beaming parameter $\eta$ vs. phase angle for cases in which the beaming parameter was actively fit.  Black dots are NEOs, cyan dots are Main Belt asteroids, and green dots are Trojan asteroids.  The weighted least squares fit line to the WISE NEO observations (red solid line) is shown compared with that of \citet{Wolters} (blue dashed line).  Middle Left: The beaming parameter $\eta$ vs. heliocentric distance. A correlation between $\eta$ and heliocentric distance can be observed, as shown by the red line that represents a running median of all the objects including NEAs, Mars crossers, and MBAs. Middle Right: $\eta$ vs. subsolar temperature. Bottom Left: $\eta$ vs. \pv.  Bottom Right: $\eta$ vs. diameter.}
\end{figure}

For objects with measurements in at least two or more WISE bands dominated by thermal emission, the beaming parameter $\eta$ was determined using a least squares minimization but was constrained to be less than the upper bound set by the FRM case ($\pi$). Figure \ref{fig:beaming} shows the fitted $\eta$ values for objects which had measurements in two or more WISE thermal bands, along with a best-fitting double Gaussian distribution.  The median value of the 313 NEOs that had fitted $\eta$ was 1.40$\pm$0.5.  The beaming parameter could not be fitted for objects which had observations in only a single WISE thermal band; these objects were assigned $\eta=1.4\pm0.5$. Figure \ref{fig:beaming} also shows fitted beaming as a function of phase ($\alpha$) along with the best-fitting linear fit, as well as $\eta$ as a function of heliocentric distance and subsolar temperature ($T_{ss}$).  The weighted best-fit relationship between $\eta$ and phase $\alpha$ is given by $\eta = (0.00963\pm0.00015)\alpha + 0.761\pm0.009$.  Figure \ref{fig:beaming} shows that $\eta$ is strongly correlated with the heliocentric distance at which a given object was observed as shown by the red line representing the running median for the NEOs, Mars crossers and Main Belt asteroids; the closer an object is to the Sun, the higher its $\eta$ value for objects observed near 90$^\circ$ elongation.  Because NEATM assumes that an asteroid's night side contributes essentially no thermal flux, objects that are observed at high phase angles (e.g. far away from the subsolar point) may have $\eta$ values that vary in model-dependent ways.  Since WISE always observed near 90$^\circ$ solar elongation, NEOs in particular tended to be observed at high phase angles, meaning that part of the night sides were observed.  If the nightside temperature is not zero as assumed by NEATM, the relationship between $\eta$ and heliocentric distance/phase angle can vary depending on the real temperature distribution.  While the $\eta$ parameter can serve as an indicator of thermal inertia at low to moderate phase angles, this result suggests that attempting to use $\eta$ as a gauge of thermal inertia or regolith properties at high phase angles must be done with caution, due to the phase-angel dependence of $\eta$.  In general, detailed thermophysical modeling should be used if possible to unravel the various thermal and geometric effects contributing to the observed thermal continuum.

\begin{figure}
\figurenum{8}
\includegraphics[width=6in]{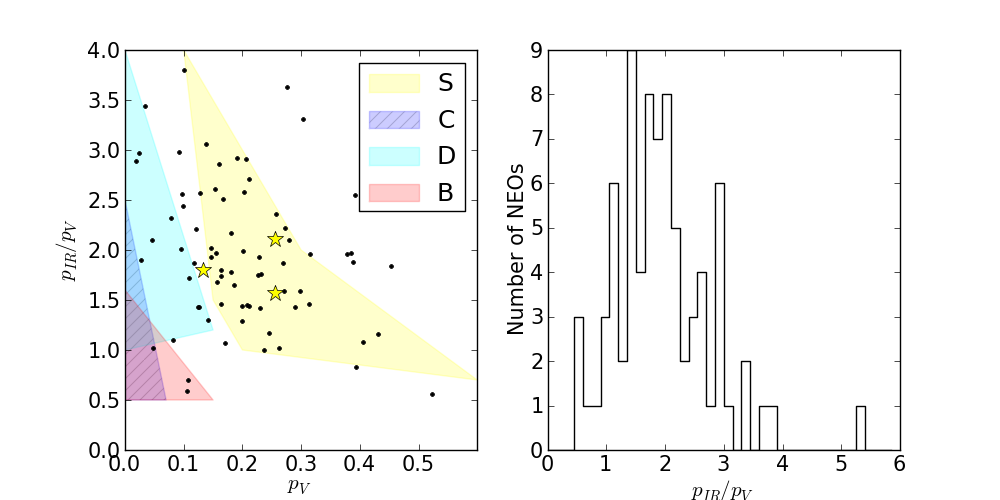}
\caption{\label{fig:pIRpV}Left: As discussed in \citet{Mainzer11d}, \irfactor\ can be used to distinguish objects with different taxonomic classifications that otherwise have similar \pv; objects with blue or neutral visible and near-infrared (VNIR) spectral slopes such as C and B types tend to have lower \irfactor\ values than objects with red VNIR slopes, such as T and D types.  NEAs with low \irfactor\ and low \pv\ are likely to be C or B types, and objects with high \irfactor\ but low \pv\ are more likely to be D types.  The shaded regions represent the regions covered by various spectroscopically classified Main Belt asteroids shown in \citet{Mainzer11d}. NEOs with known S-type spectroscopic classifications are shown as large yellow stars. Right: Histogram of \irfactor\ values for NEOs observed during the cryogenic portion of the mission. }
\end{figure}

As described above, bands $W1$ and $W2$ consist of a mix of reflected sunlight and thermal emission for NEOs, and bands $W3$ and $W4$ consist purely of thermal emission.  In order to properly model the fraction of total emission due to reflected sunlight in each band, it was necessary to determine the infrared albedo $p_{IR}$. We have made the assumption that $p_{3.6\mu m} = p_{4.6\mu m} = p_{IR}$, although it is known that there are a number of material absorption features at these wavelengths (see \citet{Gaffey02} for an overview; the topic is treated in more detail in \citet[][hereafter M11D]{Mainzer11d}).  In order to test the validity of the assumption that the infrared albedo is the same in bands $W1$ and $W2$, we recomputed all the thermal model fits with band $W1$ only and found that the resulting diameters, visible albedos and infrared albedos were identical to within $\pm$10\% 1-$\sigma$. The geometric albedo $p_{V}$ is defined as the ratio of the brightness of an object observed at zero phase angle to that of a perfectly diffusing Lambertian disk of the same radius located at the same distance.  The Bond albedo ($A$) is related to the visible geometric albedo $p_{V}$ by $A\approx A_{V} = qp_{V}$, where the phase integral $q$ is defined such that $q=2\int \Phi(\alpha) sin(\alpha) d\alpha$. $\Phi$ is the phase curve, and $q=1$ for $\Phi=max(0,cos(\alpha))$. $G$ is the slope parameter that describes the shape of the phase curve in the $H-G$ model of \citet{Bowell} that describes the relationship between an asteroid's brightness and the solar phase angle.  For $G=0.15$, $q=0.384$.   We make the assumption that $p_{IR}$ obeys the same relationship, although it is possible it varies with wavelength, so what we denote here as $p_{IR}$ for convenience may not be exactly analogous to $p_{V}$.  We can derive $p_{IR}$ directly from the WISE objects that have observed reflected sunlight in bands $W1$ and $W2$ as well as observations in $W3$ or $W4$. Figure \ref{fig:pIRpV} shows the ratio of $p_{IR}$ vs. $p_{V}$ for the objects that had fitted $p_{IR}$.  For the NEOs for which $p_{IR}$ could not be fitted, we used $p_{IR}/p_{V}=1.6\pm1.0$. 

As discussed inM11D and \citet[][hereafter M11E]{Mainzer11e}, asteroids with different taxonomic classes can have different \irfactor\ values despite having similar \pv; Figure \ref{fig:pIRpV} shows the comparison between \pv\ and \irfactor\ for the WMOPS-detected NEOs for which \irfactor\ could be fitted.  For example, a correlation was found between spectral classes that have neutral or blue visible/near-infrared (VNIR) spectral types, such as C or B types, and objects that have lower values of \irfactor, even though these spectral classes have nearly identical \pv.  The figure shows the areas where most of the major taxonomic types group, although it should be noted that most of the objects studied in M11D and M11E were Main Belt asteroids.  The extent to which the same groupings in \pv\ and \irfactor\ apply to NEOs is yet to be determined.  Taxonomic types with red VNIR slopes such as T or D types, tend to have higher values of \irfactor. It is possible that the NEOs shown in Figure \ref{fig:pIRpV} with low \pv\ and low \irfactor\ are C or B types, whereas those with high \irfactor\ and low \pv\ are more likely to be T or D types.  As discussed in M11D, the current set of spectroscopically classified objects is heavily affected by selection biases; adding more objects drawn from the NEOWISE sample would improve our understanding of the correspondence between VNIR spectroscopic shapes and features with size, \pv, and $p_{IR}$.  
 
 \subsection{Thermal Model Error Analysis}
Error bars on $D$, $p_{V}$, $p_{IR}$, $\eta$ and $T_{ss}$ were determined for each object by running 50 Monte Carlo trials that varied the objects' $H$ values by the errors described above and the WISE magnitudes by their error bars using Gaussian probability distributions.  The minimum magnitude error for all WISE measurements was taken to be 0.03 mag per the in-band repeatability given in \citet{Wright}, unless the sources were brighter than the limits $W1=6$, $W2=6$, $W3=4$, and $W4=0$; these objects were assumed to have magnitude errors of 0.2 mag in their respective bands.  The error bar for each object's model magnitude was equal to the weighted standard deviation of all the Monte Carlo trial values.  For objects with fixed $\eta$, errors on derived parameters were computed by varying $\eta$ by 0.5; this is approximate width of the Gaussian that was fitted to the beaming parameter for objects with fitted beaming (Figure \ref{fig:beaming}). For objects for which $p_{IR}$ could not be fitted, the Monte Carlo trials varied \irfactor\ by 1.0.

As described in M11B and \citet{Mainzer11c}, the minimum diameter error that can be achieved using WISE observations is $\sim10\%$, and the minimum relative albedo error is $\sim20\%$ for objects with more than one WISE thermal band for which $\eta$ can be fitted.   Table 1 gives the results of the thermal model fits for the 428 NEOs detected by WMOPS during the fully cryogenic NEOWISE mission using the WISE First Pass data processing pipeline. 

\subsection{Interesting NEOs}
\subsubsection{Large Amplitude Brightness Variations}
\begin{figure}
\figurenum{9}
\plottwo{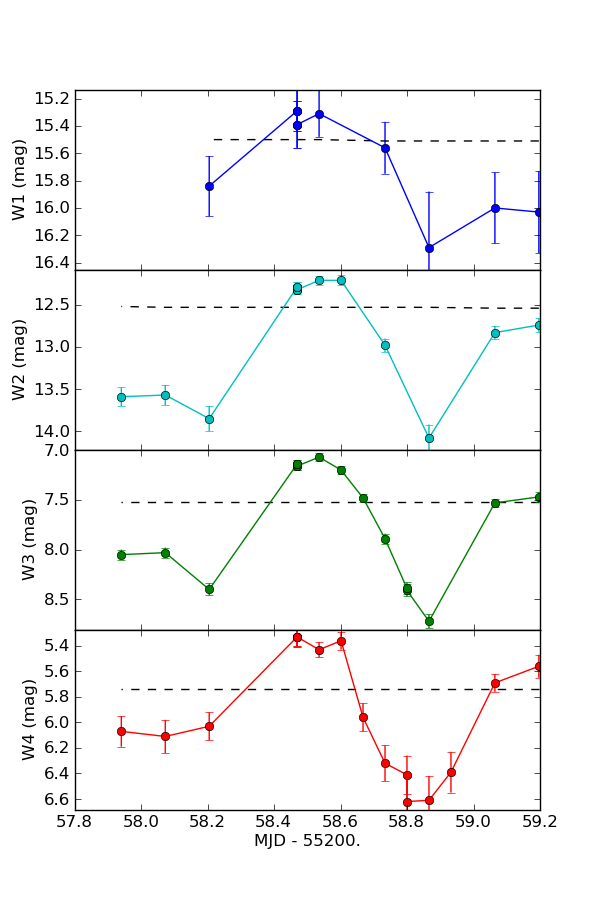}{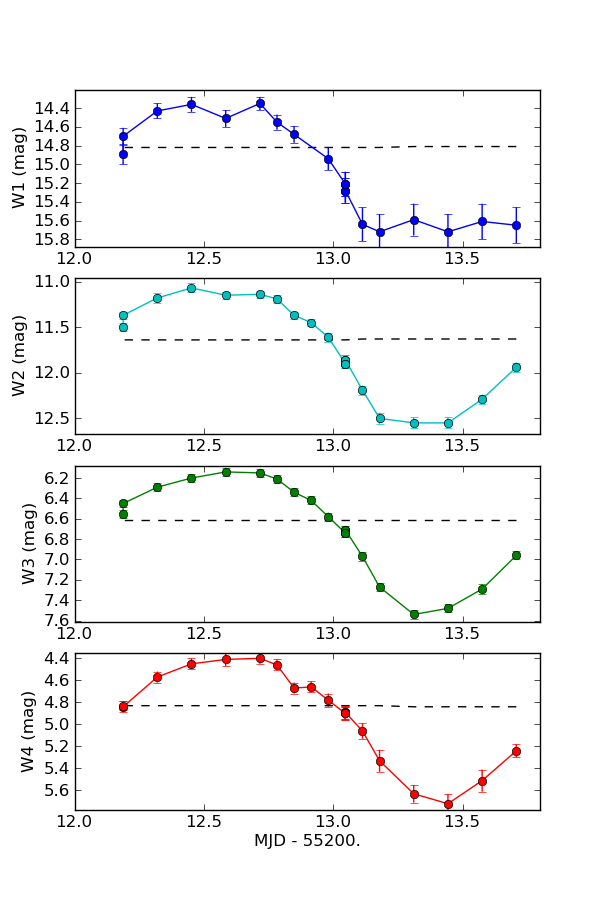}
\caption{\label{fig:big_lightcurves}Left: An example of an NEO with a large peak-to-peak lightcurve variation, (5645).  This object has a peak-to-peak amplitude of $1.29\pm0.08$, $1.65\pm0.06$, and $1.87\pm0.08$ magnitudes in bands $W2$, $W3$ and $W4$, respectively.  (5645) is known to have a 30.39 hour rotational period measured from visible photometry and is likely to be tumbling \citep{Warner}.  A future work will convolve the visible lightcurves of this and other objects with the NEOWISE infrared detections. Right: NEO 2009 WO6 also has a large-amplitude magnitude variation that is observed in all four WISE wavelengths. }
\end{figure}

A number of NEOs stood out as objects of interest.  As discussed above, we have identified a number of objects with large-amplitude peak-to-peak variations.  Figure \ref{fig:big_lightcurves} shows examples of NEOs with very large amplitude magnitude variations, indicating that these objects are likely to be highly elongated.  WISE preferentially observed NEOs at higher phase angles, which may skew the observed amplitude.  Effective diameters derived from the application of a spherical NEATM model to such objects must be viewed with caution.  These objects would benefit both from obtaining additional observations in visible light to better constrain their shapes and rotational states as well as modeling them as non-spherical shapes.

\begin{figure}
\figurenum{10}
\includegraphics[width=3in]{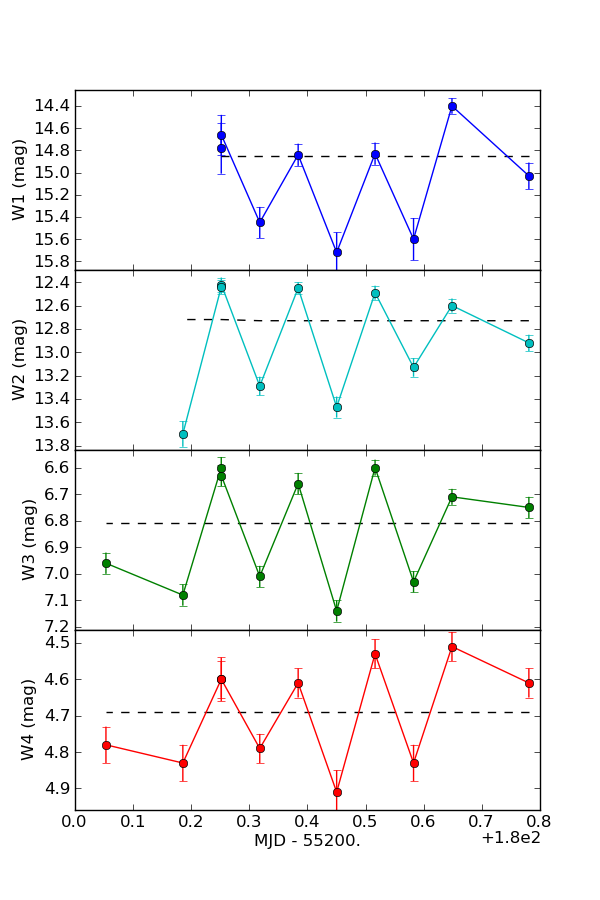}
\caption{\label{fig:1865}The NEO (1865) Cerberus has a significantly larger lightcurve amplitude at the shorter WISE wavelengths than the longer ones: $1.31\pm0.12$, $1.28\pm0.08$, $0.54\pm0.06$, $0.39\pm0.07$ in bands $W1$, $W2$, $W3$, and $W4$ respectively.  Cerberus is known to have a 6.810 hour rotational period \citep{Warner}, and its WISE-derived diameter is 1.6 km.  The high beaming parameter $\eta=2.94\pm0.03$ suggests that this could be an object with high thermal inertia; however, objects with low heliocentric distances tend to have higher $\eta$ values.  The difference in amplitudes between shorter and longer WISE wavelengths could be due to the fact that thermal emission, which dominates bands $W3$ and $W4$, can come from a larger fraction of the object's total visible area than can reflected sunlight, particularly if it has high thermal inertia and therefore a more uniform temperature distribution.}
\end{figure}

Many NEOs in the NEOWISE dataset are observed to have larger peak-to-peak amplitude variations in bands $W1$ and $W2$ as compared with $W3$ and $W4$.  This is possibly due to the fact that thermal emission can emanate from beyond the terminator on an asteroid, depending on its thermal inertia, whereas reflected sunlight tends to come from only those portions that are illuminated.  Since bands $W1$ and $W2$ consist of a mix of reflected sunlight and thermal emission, and since some NEOs have higher thermal inertias than others, a discrepancy between amplitudes at the different wavelengths can occur.  An example of such an object can be found in NEO (1865) Cerberus (Figure \ref{fig:1865}).  In addition to having significantly higher peak-to-peak lightcurve amplitudes in bands $W1$ and $W2$ than in bands $W3$ and $W4$, this object has a high beaming parameter, $\eta=2.94\pm0.03$.  This is close to the maximum $\eta$ of $\pi$ allowed by the Fast Rotating Model, which postulates that the object's cooling time is slow compared to its rotational rate; however, as shown in Figure \ref{fig:beaming} and as discussed above, $\eta$ is correlated with heliocentric distance for WISE-observed NEOs, and (1865) was located 1.09 AU from the Sun when it was detected by NEOWISE.  The object has a known rotational period of 6.810 hours \citep{Warner}, and this appears consistent with the rotational rate that can be observed in the figure.  It may be that for this object, the assumption that temperature decreases as \begin{equation}T(\theta,\phi) = T_{ss} [max(0,\cos\theta \cos\alpha + \sin\theta \sin\alpha \cos\phi)]^{1/4}\end{equation} is inappropriate, and an improved model of temperature distribution may be desirable.  NEATM assumes that $T(\theta,\phi)$ goes to zero on the night side of the asteroid, even though this may not be correct.  A number of other NEOs in our sample have high beaming parameters, and these objects are candidates for having high thermal inertia and/or rapid rotational rates. Full exploration of the true physical meaning of the relationship between $\eta$, heliocentric distance, rotational rate, and properties such as thermal inertia will be the subject of future work.

\subsubsection{Multi-Epoch Observations}
Due to the WISE observing cadence, there were 55 NEOs in our current sample during the fully cryogenic portion of the mission that were observed at two separate epochs; three NEOs were observed at three epochs. Of these, 20 have diameter measurements that agree to within 10\% between all available epochs; the remaining 35 objects' diameters agree to better than 10\% between epochs.  However, three NEOs have diameter measurements that are disparate at the 30\% or greater level.  Of these, all have $W2$ or $W3$ peak-to-peak amplitudes larger than 0.3 mag, indicating that they are likely to have elongated shapes.  Elongated objects observed at different viewing angles are likely to be poorly fit by a spherical thermal model.  Furthermore, large disparities between observing epochs can be indicative of large differences between temperature on the evening and morning sides; more sophisticated thermal models will need to be developed.  In order to compute a single effective diameter and albedo for purposes of computing the debiased population statistics, the thermal model was fit using observations from all epochs simultaneously.  These single effective diameters were used as an input to the debiasing model; nevertheless, these objects should be fit with non-spherical thermal models.  The objects with relatively low amplitude lightcurve variations that have large discrepancies in their fitted diameters and albedos between the two epochs are possibly good candidates for having significant temperature differences between their morning and evening sides. Objects with large morning/evening temperature differences could be subject to increased Yarkovsky forces.    

\section{Debiasing}
In order to determine the size and albedo distributions of the NEOs, it is necessary to remove the effects of any systematic survey biases present in the NEOWISE sample.  NEOWISE carried out a ``blind" search for moving objects, meaning that all moving objects were detected in the same way regardless of whether or not they had been previously discovered by another observer.  This means that the NEOWISE survey can be debiased independently of the biases of other surveys with the following caveats: There were 12 objects that were designated without visible follow-up and 22 objects that appeared on the NEO Confirmation Page but received neither designations nor follow-up; there is an undetermined number of tracklets that may have been misclassified by the MPC as non-NEOs.  Future work will attempt to assess the fraction of real NEOs that are misclassified. 

In order to model the NEOWISE survey biases, a high-fidelity simulation of the NEOWISE survey was created.  As described in M11A and the WISE Explanatory Supplement, the WMOPS pipeline required five independent detections of a particular object for it to have been considered detected.  Sources were extracted down to 4.5 $\sigma$ in the Level 1b images.  We used the Julian dates and coordinates of the centers of all pointings that were used by WMOPS throughout the fully cryogenic portion of the survey as well as the on-sky footprint of each pointing (47 x 47 arcminutes) to recreate the survey history.  As described in \citet{Wright}, the WISE survey includes hundreds of observations of the ecliptic poles and an average of 10-12 observations of most moving objects in the ecliptic plane.  Furthermore, due to the observational cadence described in M11A and \citet{Wright}, WMOPS is sensitive to objects with an apparent velocity between 0.06 to 3.2 $^{\circ}$/day.  The simulated survey replicated these selection criteria for identifying a moving object tracklet.

We first determined the number of NEAs with effective diameters larger than 1 km by collecting together as much information as possible about all known NEAs (whether observed by WISE or by others) with $H<22$, since that is the approximate limit at which an asteroid with \pv$=1\%$ could still have $D=1$ km. Where possible, we used diameters determined from radar observations, in situ spacecraft imaging, or radiometric measurements from WISE, IRAS, or ground-based thermal infrared observations.  With NEOWISE, we detected 175 NEOs with $D>1$ km; when we added in all the NEOs with measured diameters that we could find from the literature, we found an additional 75 objects, for a total of 250 objects with previously known effective diameters that are $>$1 km.  We also found from the literature an additional 72 NEAs with taxonomic classifications and assigned them approximate albedos based on the average albedos given in Table 1 of M11D.  We note that most of the objects studied in M11D were Main Belt asteroids, and albedos associated with a particular taxonomic class were found to be strongly affected by selection biases below $\sim$30 km.  The extent to which the albedo distributions for various taxonomic classes for the NEOs resemble those of large Main Belt asteroids is unknown, since the spectroscopic samples used in M11D and M11E are heavily biased against optically faint objects, which preferentially tend to be small, low albedo objects. Using the albedo distributions for different taxonomic classes derived from large Main Belt asteroids introduces an unknown error, and the error estimates we quote here must be regarded as lower limits.  Our estimates will be improved by finding more NEOs within the WISE dataset and directly computing diameters for them.  We used the albedo distribution of the 250 objects with known diameters to compute $D$ for the previously known NEOs with $H<22$ mag but no diameter or albedo measurements using the relationship \begin{equation}p_{v} = \left[\frac{1329\cdot 10^{-0.2H}}{D}\right]^{2}.\end{equation} Applying the albedo distribution of the $H<$22 objects with known diameters gave us a total of 911$\pm$17 previously known NEOs with diameters $>$1 km.  

Next, it was necessary to determine the remaining population of 1 km NEOs likely to exist that have not yet been discovered.  NEOWISE discovered 16 new NEOs with $D>1$ km.  The total undiscovered population remaining was found by determining the NEOWISE survey bias for objects in this size range, then debiasing the NEOWISE disocoveries of large NEOs. The survey bias was found by creating a synthetic population of large NEOs and running a simulated WMOPS survey on this population to identify what would have been found.  In order to produce a statistically valid sample, we created 25 populations each with 50,000 synthetic NEOs with orbital elements generated randomly using the synthetic solar system model \citep[S3M;][]{Grav}.  In the S3M model, the NEO orbital elements are generated according to the distribution created from five source regions given in \citet{Bottke}. Visible albedo and $p_{IR}$ were randomly assigned values between zero and one; $\eta$ was randomly assigned a value between zero and $\pi$.  Diameter was assigned a random value between 1-10 km.  The next step was to compute the survey bias by integrating the synthetic objects' orbits over the entire list of $\sim$700,000 pointings carried out during the cryogenic portion of the WISE survey and see which objects were present in each frame.  We then determined whether or not the objects would have been detected by NEOWISE by modeling their fluxes and comparing them to a model of the WISE sensitivity to moving objects in all four bands.  Several other trial diameter, $\eta$, and \pv\ distributions were used, and as expected, the survey bias is not sensitive to the input distributions used as long as there are enough synthetic objects in each bin to reduce statistical error from the simulation.  

\begin{figure}
\figurenum{11}
\includegraphics[width=6in]{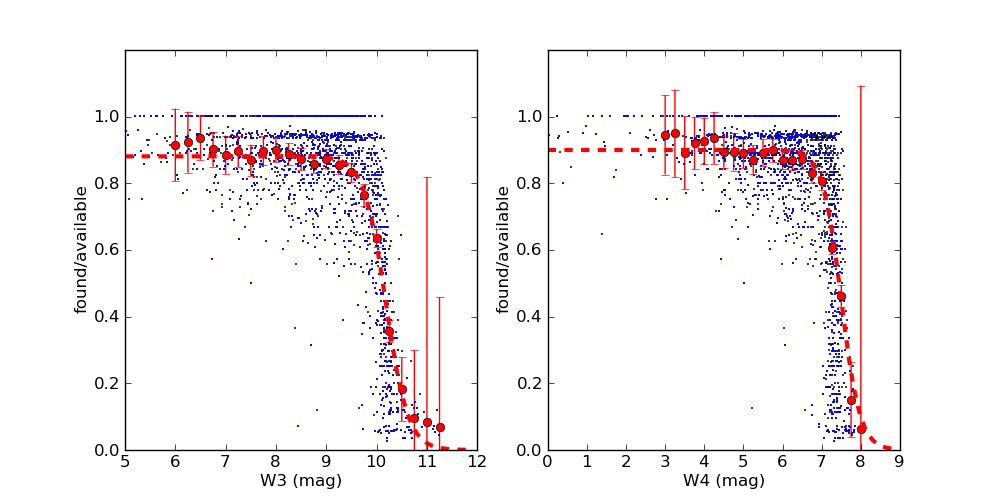}
\caption{\label{fig:ssoid} The Known Solar System Object Possible Association List was used to assess the solar system object detection probability of the NEOWISE survey during the cryogenic portion of the mission across the sky.  This figure shows an example of the detection probability for a 40$^\circ$ x 20$^\circ$ region centered at 0$^\circ$ ecliptic latitude and 0$^\circ$ ecliptic longitude.  The dots show $n_{detected}/n_{available}$ as a function of observed magnitude for individual numbered asteroids, where $n_{detected}$ is the number of times an object was actually detected out of $n_{available}$ possible times the object was in a WISE frame.  The quantization observed at 90\% and 100\% is due to the fact that on average, WISE observes most moving objects $\sim$10 times, and bright objects are usually detected 9 or 10 times.  The red circles are the medians of $n_{detected}/n_{available}$ for each magnitude bin. The function given in Equation 2 was fitted to the red circles (dashed red line); this represents the detection probability. }
\end{figure}

We used the WISE Known Solar System Object Possible Association List (KSSOPAL) to assess the survey detection efficiency to moving objects at various places across the sky.  KSSOPAL uses a list of known minor planet ephemerides to predict where asteroids should be in each WISE frame and to generate a list of probable matches; however, unlike WMOPS, it makes no attempt to eliminate matches to inertially fixed sources such as stars or galaxies, nor does it remove spurious associations with artifacts or cosmic rays.  We searched the KSSOPAL for numbered asteroids only as these generally have well-determined orbits.  In order to reduce the possibility of spurious associations with stars and galaxies, we checked each source location from KSSOPAL against the WISE Level 3 Atlas source table and used the $n$ out of $m$ statistics provided to search for sources that repeated; these sources were flagged.  For each magnitude bin, we computed the total number of available detections predicted by KSSOPAL and compared this to the total number of matches found.  This result, shown in Figure \ref{fig:ssoid}, gives an estimate of the single image completeness as a function of flux for a particular region of the sky for bands $W3$ and $W4$.  We computed this completeness curve for a number of different locations throughout the sky to bracket the WISE survey sensitivity as a function of ecliptic latitude/longitude and distance from the Galactic Center.  Figure \ref{fig:ssoid} shows the probability that a moving object of a particular flux was detected by the WISE pipeline at a particular location on the sky.  Based on the results from KSSOPAL, we found that the detection probability dropped to zero within five degrees of the Galactic Center.  The detection probability curves $P$ were fitted with the following function for both bands $W3$ and $W4$: \begin{equation}P = \frac{a_{0}}{2}\left[ 1- \tanh{\left( a_{2} (W - a_{1}) \right)}\right] \end{equation} where $W$ is the $W3$ or $W4$ magnitude and $a_{i}$ are the fitted coefficients.  For sources between $5 - 25 ^{\circ}$ of the Galactic Center, the detection probability coefficients were given by $a_{0}=0.9$, $a_{1}=9.5$, $a_{2}=1.0$ for band $W3$ and $a_{0}=0.9$, $a_{1}=7.0$, $a_{2}=2.0$ for band $W4$; for sources at all further distances from the Galactic Center, the coefficients were given by $a_{0}=0.9$, $a_{1}=10.25$, $a_{2}=2.5$ for band $W3$ and $a_{0}=0.93$, $a_{1}=7.5$, $a_{2}=2.1$ for band $W4$.  

\begin{figure}
\figurenum{13}
\includegraphics[width=3in]{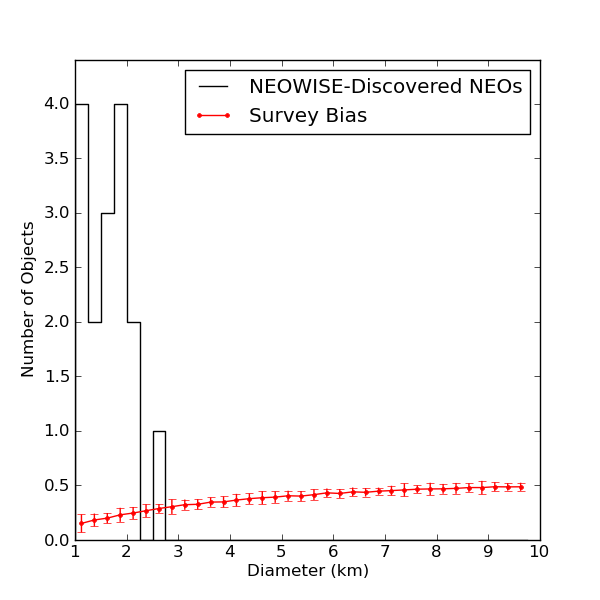}
\caption{\label{fig:discovered_neos} The NEOWISE survey bias for objects with $D>1$ km (red line with points) was computed by creating a synthetic population of NEOs with orbits generated using the orbital element distribution of \citet{Bottke}.  The number of NEOWISE-discovered NEOs $>$1 km is shown as the solid black line.  The total number of remaining undiscovered NEOs was determined by dividing the NEOWISE-discovered NEO distribution by the survey bias, then integrating the result over diameter.}
\end{figure}

The synthetic population of near-Earth asteroids with $D>1$ km was compared to the list of frames taken throughout the cryogenic portion of the survey to determine whether or not a particular object was present in the WISE field of view.  Known comets were not considered in the population at this point.  As discussed above, physical parameters were randomly assigned to each object and fluxes were computed using the NEATM model.   In order for an object to be detected by the synthetic survey, it had to meet the WMOPS selection criteria: the object had to appear five or more times and had to have a relative velocity in the range 0.06 to 3.2 $^\circ$/day.  We note that the WMOPS software is not sensitive to the fastest-moving NEOs, which will tend to be small objects that are very close to Earth.  This bias against fast-moving NEOs will result in an an additional loss of sensitivity to the smallest objects; we will attempt to model this bias in future work. The survey bias for objects with $D>1$ km was found by dividing the population of synthetic objects in each size bin by the population found by the simulated NEOWISE survey.  The survey bias in diameter for objects $>$1 km is shown in Figure \ref{fig:discovered_neos} along with the NEOWISE-discovered NEAs.  The total number of remaining undiscovered NEAs $>$1 km was computed by dividing the NEOWISE-discovered population's differential size distribution by the survey bias, then integrating the result over diameter, resulting in $\sim$80$\pm$9 NEAs remaining to be discovered in this size range.  When added to the 911$\pm$17 previously known NEAs, we compute a total of 981$\pm$19 NEAs with $D>$1 km.  This result suggests that the Spaceguard goal of detecting  90\% of all NEOs larger than 1 km \citep{Morrison} has essentially been met (although we note that our analysis does not yet address the near-Earth comets). 

\begin{figure}
\figurenum{12}
\includegraphics[width=6in]{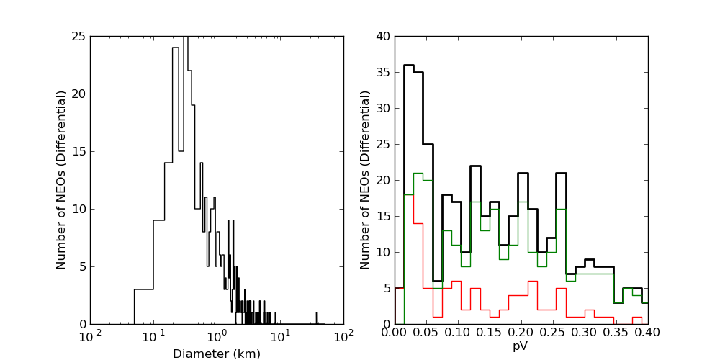}
\caption{\label{fig:differential}Left: Preliminary observed undebiased differential diameter distribution for the NEOs observed during the cryogenic portion of the mission. The peak of the distribution occurs at $\sim$300 m.  Right:  The albedo distribution for all NEOs observed by NEOWISE during the cryogenic portion of the mission are shown as the heavy black line; previously known NEOs are shown in green, and NEOs discovered by NEOWISE are shown in red.  There are proportionately more low-albedo NEOs that were discovered by NEOWISE than in the previously known population; this is evidence of the bias that visible surveys suffer against low-albedo objects. }
\end{figure}

\begin{figure}
\figurenum{14}
\includegraphics[width=6in]{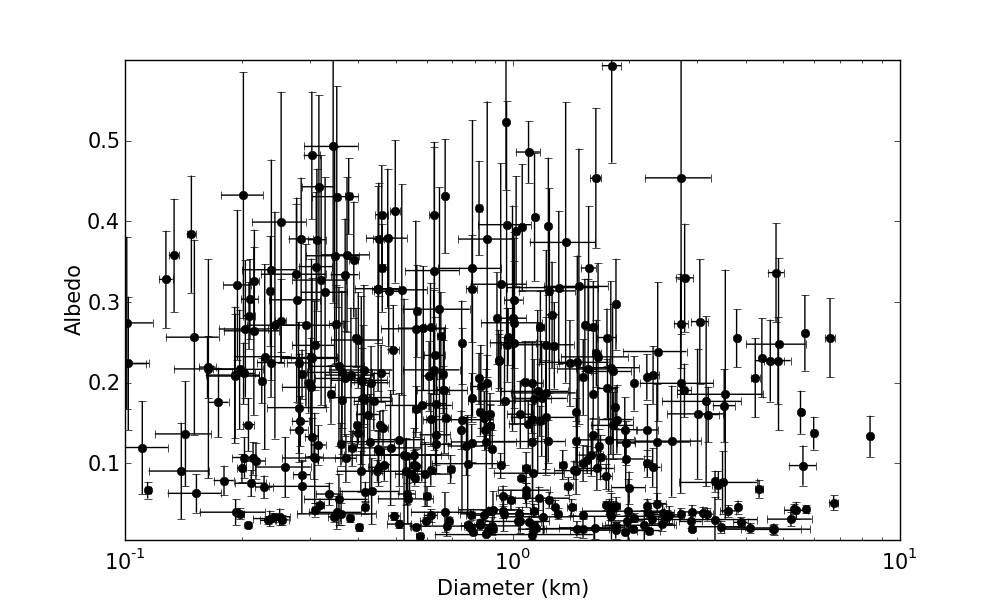}
\caption{\label{fig:diam_alb} We do not observe a strong correlation of \pv\ with diameter for the objects detected during the cryogenic portion of the NEOWISE survey.  Missing from the plot are the 12 NEOs that received designations but received no visible light follow-up, along with 22 candidate NEOs that appeared on the NEO Confirmation Page but received neither designations nor visible follow-up.  These objects are likely to be both small and dark.}
\end{figure}

We do not find strong evidence for a correlation between diameter and \pv\ from the NEOWISE dataset.  \citet{HarrisACM} and \citet{Delbo} show an apparent correlation between size, \pv, and asteroid taxonomic type. In M11D, we show that \pv\ is correlated with diameter within a particular taxonomic type, but this correlation could be partially or entirely due to the observational biases of visible light surveys and spectroscopic measurements, which are less likely to detect and characterize small asteroids with lower albedos.  Figure \ref{fig:diam_alb} shows a plot of diameters vs. \pv\ for all the NEOs observed during the cryogenic portion of the NEOWISE survey.  No strong correlation can be observed.  Furthermore, it should be noted that there were 12 NEOs that received designations from the MPC but did not have visible light follow-up, so \pv\ could not be computed for these objects.  They tended to be the most difficult for visible light observers to follow-up, implying that they are likely to have low albedos.  An additional 22 candidate NEOs appeared on the NEOCP but were not designated due to their short observational arcs, so neither diameters nor \pv\ could be computed for these objects.  Like the 12 that were designated, these objects were the most difficult for observers to detect (however, it is possible that some of these undesignated objects are Main Belt asteroids or Trojans instead of NEOs).  Finally, as discussed above, WMOPS is less sensitive to the fastest moving NEOs, which are likely to be small objects very close to the Earth, resulting in a further bias against the smallest objects.  These fast-moving small NEOs are also likely to be subject to additional bias if they have low albedos, since we require visible light follow-up to obtain an albedo measurement.  The 12 designated NEOs for which \pv\ could not be determined represent only $\sim$5\% of the 460 objects detected during the cryogenic portion of the survey, so they can only have a commensurately small effect on the error in the population statistics.  A more detailed treatment of these objects, the 22 without designations, and the bias against fast-moving NEOs will be dealt with in a future work.  For now, the error estimates presented herein must be regarded as a lower limit, and with these caveats, we conclude that there is no strong dependence of \pv\ on diameter. The \pv\ distribution shown in Figure \ref{fig:differential} was assumed to hold down to the smallest sizes.  While a rigorous comparison of our results with those of earlier studies \citep[][e.g.]{Delbo,HarrisACM} would require data for NEOs of the same taxonomic types in the same size range, our result suggests that the apparent correlations between size and albedo reported by others could be entirely due to observational biases and small sample sizes as was also shown in \citet{Mainzer11d}.  Linkages between albedo, size and space weathering cannot be reliably made without accounting for these biases.

With the total number of 1 km NEAs in hand as well as the observed distributions for \pv, \irfactor, and $\eta$, it was possible to compute the total numbers of objects at smaller size ranges.  As before, the total number of objects with diameters $>$100 m was determined by computing the survey diameter bias in each size bin and dividing it into the NEOWISE-observed diameter distribution. As was done for the $>$1 km NEAs, the survey biases down to 100 m were determined by creating a synthetic population of objects and computing which synthetic objects would have been detected by WMOPS.  Instead of creating synthetic objects with flat distributions for \pv, \irfactor, and $\eta$, we used the distributions for the objects that were observed by the NEOWISE survey (Figures \ref{fig:differential}, \ref{fig:pIRpV}, and \ref{fig:beaming}).  The visible albedo was assigned using a probability function described by a double Gaussian: \begin{equation}P(p_{V})=v_{0}e^{-(p_{V}-v_{1})/2v_{2}^2}+v_{3}e^{-(p_{V}-v_{4})/2v_{5}^2}\end{equation} where $v_{0}=12.63$, $v_{1}=0.034$, $v_{2}=0.014$, $v_{3}=3.99$, $v_{4}=0.151$, and $v_{5}=0.122$. The beaming parameter $\eta$ was chosen using a double Gaussian probability distribution, and \irfactor\ was generated using a single Gaussian probability distribution.  

The cumulative diameter distribution was initially modeled as a broken power law ($N>D^{-\alpha}$) with $\alpha=5$ above 5 km and $\alpha=2.1$ for $D < 5$ km based on \citet{Jedicke}. Fluxes were generated for each synthetic object using the assigned physical parameters and orbits as inputs to the faceted NEATM model described above.  The broken power law was found to be a poor match to the observed cumulative diameter distribution, producing too many NEAs at the smallest sizes.  The cumulative size distribution was found to be best represented by a triple power law with breaks at 5 km and 1.5 km.  With this functional form of the cumulative size distribution, we found the slope of the power law below 1.5 km by computing the total number of NEAs with diameters $>$100 m as described above by dividing the observed differential size distribution by the survey bias, then integrating the result over diameter. Using this method, we computed a total of 20,500$\pm$3000 NEAs with $D>$100 m.  Using this number and the number of 1 km NEAs results in a slope of the cumulative size distribution power law of 1.32$\pm$0.14.  This power law slope produces $\sim13,200\pm$1,900 NEAs with $D>$140 m, the limit of the George E. Brown Congressional mandate, and $\sim51,300\pm11,500$ NEAs with $D>$50 m.   This result is lower than was predicted by previous analyses such as \citet{Harris08}, \citet{Jedicke} and \citet{Rabinowitz00}.  The break in the power law at 1.5 km is similar to that shown in \citet{Harris08}.  \citet{Harris08} shows another break in the cumulative size distribution at $D\sim$50-100 m and finds that the slope of the size distribution steepens, indicating that there are larger numbers of smaller objects in this size range.  However, as our current sample only includes four NEAs with $D<$100 m, our determination of the numbers of objects or the slope of the size distribution below 50-100 m is not reliable.  

\begin{figure}
\figurenum{15}
\includegraphics[width=6in]{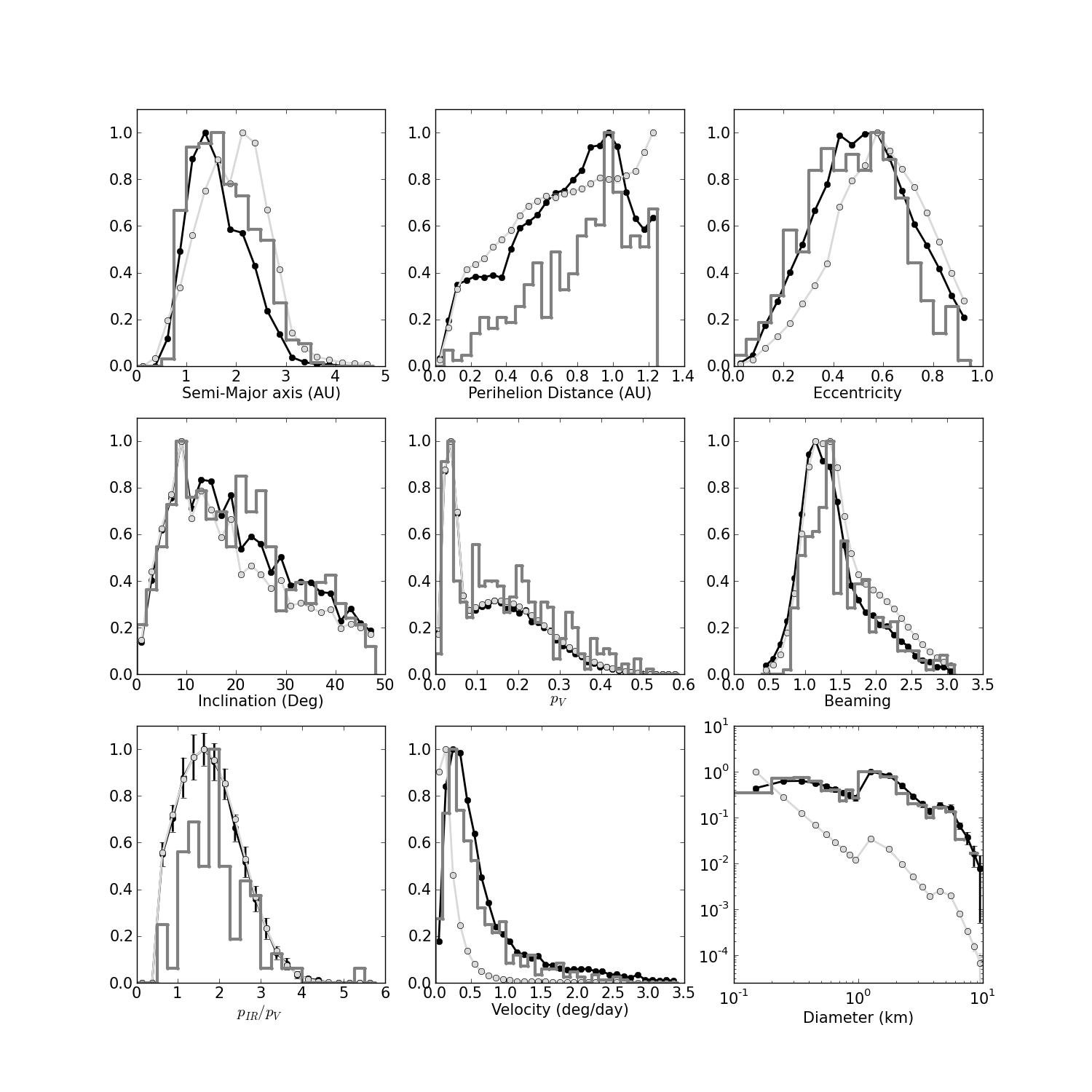}
\caption{\label{fig:synthetic_pop} This figure shows the properties of the synthetic population (light gray line and points) compared with the simulated objects detected by the simulated WMOPS survey (black line and points).  The medium gray line represents the actual objects detected by NEOWISE during the fully cryogenic portion of the survey.  All three lines represent normalized numbers of objects (hence the maximum value on the y axis is one).  To first order, the orbital elements of the synthetic population found by the simulated WMOPS survey (the ``found" objects) match the properties of the objects detected by NEOWISE (the ``observed" objects); this result indicates that the simulated objects' orbital elements are probably a good representation of the real population of NEAs. The good match between the ``found" objects and the ``observed" objects in \pv, $\eta$, and \irfactor\ is due to the fact that the simulated objects' physical properties were created to match those of the population detected by WMOPS. If we divide the synthetic population (light gray lines and points) into the ``found" objects (black lines), the results are the survey biases for each of the various parameters, which are shown as the light gray lines in Figure \ref{fig:bias_and_observed}.}
\end{figure}

\begin{figure}
\figurenum{16}
\includegraphics[width=6in]{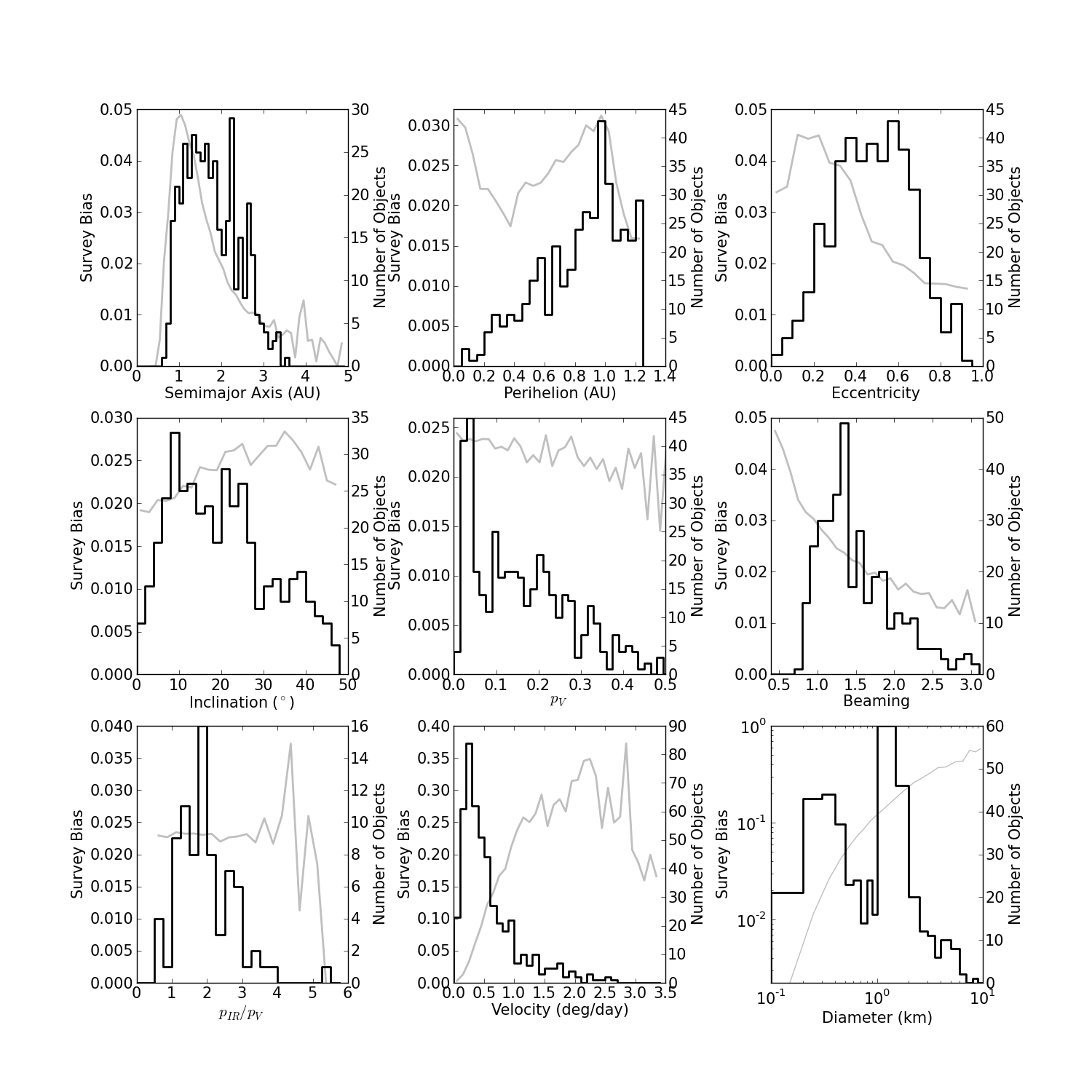}
\caption{\label{fig:bias_and_observed} This figure shows the survey biases computed by dividing the synthetic population into the simulated objects found detected by the simulated WMOPS survey (gray lines).  The black lines show the properties of the objects detected by WMOPS. It can be seen that the WISE survey is essentially unbiased with respect to visible albedo. }
\end{figure}

The survey biases in \pv, \irfactor, $\eta$, semi-major axis, eccentricity, inclination, and perihelion were determined by dividing the total synthetic population into the distributions of the objects that would have been found by the NEOWISE survey (Figure \ref{fig:bias_and_observed}).  The results show that the NEOWISE survey is essentially unbiased with respect to \pv; this is expected given the weak dependence of thermal flux on \pv.  Because the WISE survey covered the entire sky, including the ecliptic poles, the survey has a moderate bias in favor of high inclination objects.  Comparison of the synthetic objects' orbital elements produced by the \citet{Bottke} model to the observed orbital element distributions reveals that this model is reasonably consistent with the observed population (Figure \ref{fig:synthetic_pop}).  Future work will compare the characteristics of the NEOs to their probable source regions within the Main Belt and comets and will attempt to further refine the distribution of NEO orbital elements.

We finally note that it is possible that some short-arc NEOs submitted to the Minor Planet Center can be classified as Main Belt asteroids.  It is possible to ascertain the fraction of objects that are misclassified by generating and sending a set of synthetic NEOs to the MPC in the form of tracklets (right ascensions, declinations, times, and spacecraft velocities).  The digest score used to determine the probability that an object is an NEO could then be run on each synthetic tracklet to determine the fraction that would be erroneously classified as MBAs.  This analysis will be the subject of future work in order to determine the fraction of misclassified NEOs likely to exist.  At present, this fraction remains unquantified, and as stated above, the error estimates provided should be regarded as lower limits.

\subsection{NEOs As Future Targets for Exploration}
\begin{figure}
\figurenum{18}
\includegraphics[width=6in]{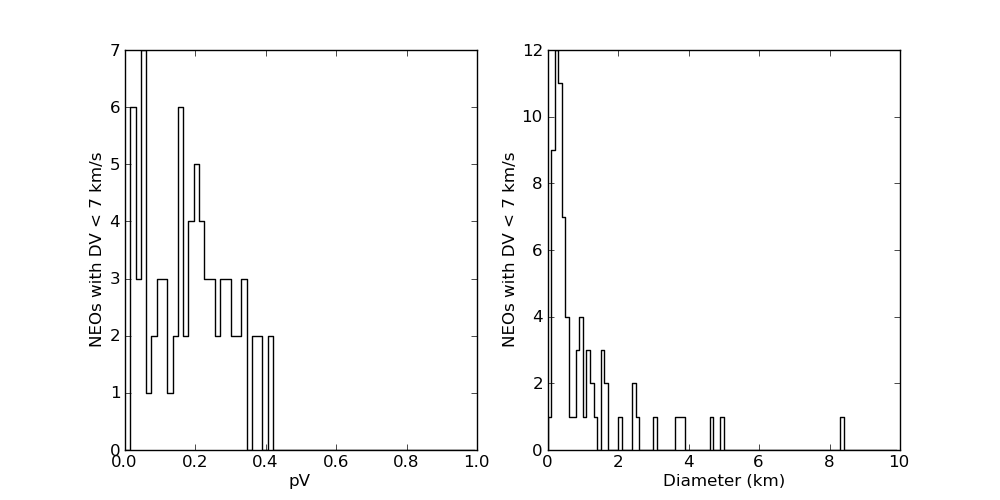}
\caption{\label{fig:Low_DV} There are 80 NEOs that were observed by NEOWISE during the fully cryogenic portion of the mission that have $\Delta v < 7$ km/s and observational arcs $>$0.1 year (observational arc gives an indication of the quality of knowledge of an object's orbit), and they span a range of albedos similar to that observed in the general NEO population by NEOWISE.  However, in practice, very few of these are suitable for human exploration, as they rarely have sufficiently close approaches in the timeframe of interest (2020-2050).  }
\end{figure}

NEOs have been discussed as possible destinations for human exploration \citep{Abell}, and there has been a recent push to obtain characterization information for objects with low $\Delta v$ \citep{ShoemakerHelin}, since these tend to require the least energy to reach.  Objects with low $\Delta v$ are easier to reach with robotic missions as well.  Figure \ref{fig:Low_DV} shows the albedo and size distributions for 80 NEOs that were observed by NEOWISE during the fully cryogenic portion of the mission that have $\Delta v < 7$ km/s as computed by Lance Benner's list of $\Delta v$ for all NEOs observational arcs longer than 0.1 years (this being a rough measure for how well an object's orbit is likely to be known; objects with observational arcs shorter than this will most likely be extremely difficult to locate in the future).  Using a similar approach, \citet{Mueller} studied 65 NEOs with $\Delta v < $7 km/s; however, while there appear to be many targets, in practice, only a handful of NEOs will actually be suitably located in the timeframe of interest for human exploration (between $\sim$2020-2050).  Among our detections of low $\Delta v$ NEOs,  (3361) Orpheus and (207945) have close approaches in the next two decades; both have moderate to high albedos ($0.28\pm0.09$ and $0.16\pm0.02$, respectively) consistent with other S-complex objects observed by NEOWISE (M11D). NEOs with low albedos are potentially targets of greater interest, as these may possess volatile materials such as water that could be used as in situ resources for explorers.  While one object among our sample (1996 GQ) has $p_{V}=0.02\pm0.002$ and $\Delta v\sim 6.5$ km/s, it may not approach Earth sufficiently closely to be readily accessible to human explorers in the 2025 timeframe.     

Many of the NEOs being considered for human exploration have very large $H$ values, ranging as high as $H=27-28$ magnitudes \citep{Abell}.  As we have shown in earlier sections, NEO albedos range widely, from $\sim$0.01 to $\sim$0.6 or higher.  If only visible photometry is available, a proposed target with $H=27$ could be as small as only 5 m.  Such an object would be similar in size (or smaller) than a visiting crew capsule: a less-than-ideal target.  It is important to provid solid diameter estimates for any potential target rather than relying solely upon $H$ due their widely varying albedos.  By virtue of observing near 90$^\circ$ solar elongation, NEOWISE was able to discover an unusual NEO during the post-cryogenic mission that may represent the first of a new class of objects well-suited for human exploration.  2010 TK7 \citep{Connors} is the first known Earth Trojan asteroid; although its inclination is too high to make its relative velocity low enough to be easily accessible, similar objects in more energetically favorable Trojan orbits may exist.  

\section{Conclusions}
The NEOWISE project has resulted in the acquisition of a sample of NEOs that is essentially unbiased with respect to visible albedo, allowing us to compute the numbers, sizes, and albedos of NEAs down to $\sim$100 m with reduced errors relative to previous work.  By virtue of being in space, NEOWISE is characterized by its well-known sensitivity, number and location of pointings, and consistent image quality.  We have shown that the Spaceguard goal of identifying 90\% of all NEAs larger than 1 km has been met and exceeded.  Furthermore, the number of NEAs with diameters as small as 100 m is likely to be less than previously suspected. We must apply similar analysis techniques to the subset of the NEAs that are potentially hazardous to determine whether this implies that the hazard is commensurately lower; this will be the subject of future work.  We note that this analysis applies only to near-Earth asteroids; a future work will assess the population of near-Earth comets, as their variable levels of activity can complicate efforts to obtain sizes and albedos for them \citep[c.f.][]{Bauer11a,Bauer11b}.  

Using the methods described above, we find that of the $\sim$8,000 NEOs known to date, $\sim$5200 of them are larger than 100 m.  Since we estimate that there are 20,500$\pm$3000 NEAs in total with $D>$100 m, the $\sim$5,200 that have been found to date represent a relatively small fraction of the total that exist, implying that many still remain to be discovered.  However, our sample contains only handful of objects smaller than 100 m, so we are unable to confirm whether or not the cumulative size distribution slope remains the same below this limit.  Although \citet{Harris08} found that the size distribution slope breaks at $D\sim$50-100 m and has a steeper slope below this limit, suggesting that the relative number of smaller objects is larger, we are unable to comment reliably on the numbers of NEAs below this size range with the existing NEOWISE sample at present.     

Contrary to previous analyses with smaller datasets, we find no strong correlation of size with visible albedo.  This result suggests that previous work (including those analyses attempting to link size and albedo to space weathering) has been hampered by observational biases, although the current lack of taxonomic classifications of most of the small NEOs observed by NEOWISE precludes a firm conclusion on this point \citep[for a more extensive discussion of the links between size, albedo and taxonomic classifications, see also ][]{Mainzer11d}.  We note that the inclusion of the relatively small fraction of NEOWISE-discovered NEOs that did not receive visible follow-up is likely to increase the numbers of small, low albedo objects, further reducing the likelihood of a correlation between albedo and size.  Compared with the average distribution of visible albedos found in the Main Belt \citep{Masiero}, the NEOs are preferentially brighter, following a roughly bimodal distribution with $\sim$40\% having \pv$<$0.1.  As expected given that we have shown that WISE is essentially unbiased with respect to \pv, NEOWISE preferentially discovered a higher fraction of low albedo NEOs than that found by visible light surveys, with 53\% having \pv$<$0.1.  With the improved number, size and albedo distributions of the NEAs in hand, we can now begin to refine our understanding of their probable source regions within the Main Belt and the comets.   

Our thermal models have revealed that the NEATM beaming parameter $\eta$ is correlated with phase angle/heliocentric distance, resulting from the fact that the temperature distribution as a function of latitude and longitude across an asteroid's surface is more complex than that assumed in the NEATM.  Nevertheless, we have identified a number of NEOs with unusual properties such as (1865) Cerberus and NEOs with large-amplitude lightcurves; our understanding of these objects would benefit from continued study and additional follow-up observations.  Similarly, the assumption that all objects can be reasonably represented as spherical should be revisited, and attempts should be made to determine three-dimensional shape models and rotational states.  

We note that our results will be improved by the incorporation of the second generation of WISE data processing, as well by continued visible light follow-up of NEOWISE-discovered NEOs, both to improve their orbits and to refine their $H$ and $G$ values. The albedos that we have computed depend entirely on the quality of the $H$ and $G$ values that underpin them; while we have accounted for typical random errors associated with $H$ and $G$, if these parameters have systematic offsets or trends, these effects can change the computed albedos.  Our understanding of the numbers, sizes, and albedos of the NEOs will be improved by detection of more objects in the WISE dataset at lower signal-to-noise values as well as by mining the WISE data in order to compute diameters and albedos for previously known objects that can only be detected by combining all available images of them.  We must also determine the debiased population statistics of the near-Earth comets and potentially hazardous objects in order to better constrain the characteristics of these populations; this will be the subject of future work.  The fraction of NEOs that are not identified correctly by the MPC digest score remains unquantified at present.  Nevertheless, even with these caveats, the NEOWISE portion of the WISE project has significantly improved our understanding of the NEAs and paves the way for future studies.

\section{Acknowledgments}

\acknowledgments{This publication makes use of data products from the \emph{Wide-field Infrared Survey Explorer}, which is a joint project of the University of California, Los Angeles, and the Jet Propulsion Laboratory/California Institute of Technology, funded by the National Aeronautics and Space Administration.  This publication also makes use of data products from NEOWISE, which is a project of the Jet Propulsion Laboratory/California Institute of Technology, funded by the Planetary Science Division of the National Aeronautics and Space Administration.  We thank our referee, Dr. Alan Harris of DLR, for his thoughtful comments which materially improved this work. We also thank Dr. Alan Harris of the Space Sciences Institute for useful converations.  We gratefully acknowledge the extraordinary services specific to NEOWISE contributed by the International Astronomical Union's Minor Planet Center, operated by the Harvard-Smithsonian Center for Astrophysics, and the Central Bureau for Astronomical Telegrams, operated by Harvard University.  We also thank the worldwide community of dedicated amateur and professional astronomers devoted to minor planet follow-up observations. This research has made use of the NASA/IPAC Infrared Science Archive, which is operated by the Jet Propulsion Laboratory, California Institute of Technology, under contract with the National Aeronautics and Space Administration.}

 \clearpage
 
\clearpage

\begin{figure}
\figurenum{17}
\includegraphics[width=6in]{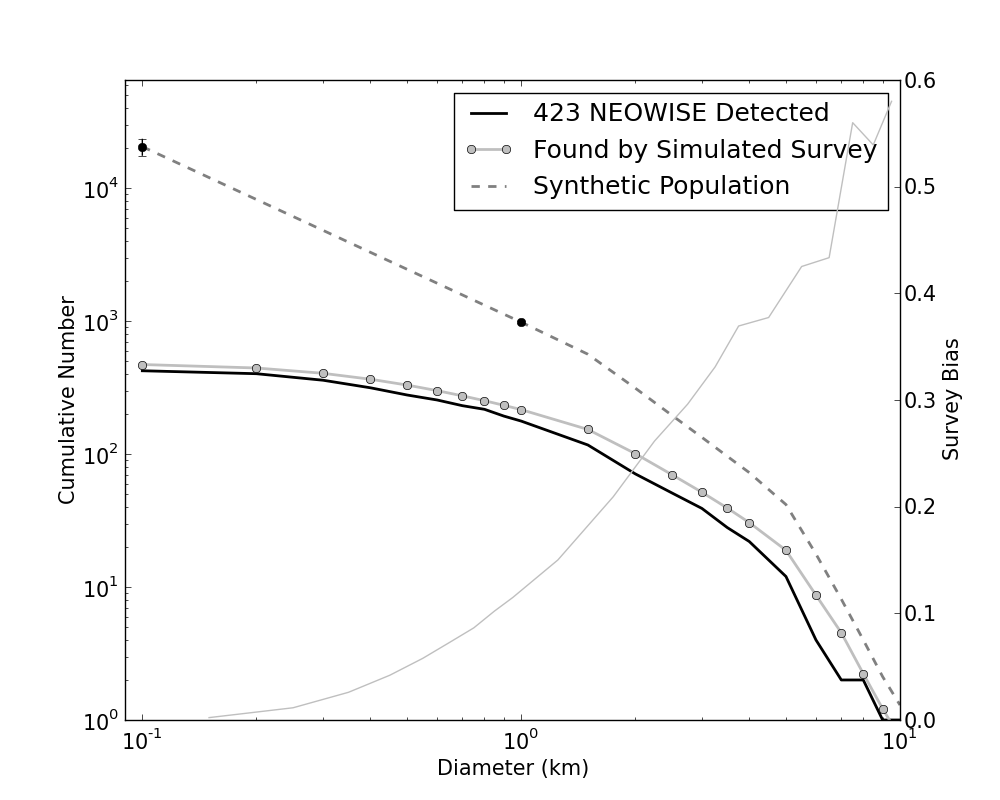}
\caption{\label{fig:cumulative_diameter} The cumulative size distribution of the synthetic population, the synthetic objects found by the simulated WMOPS survey, and the objects detected by NEOWISE.  This synthetic population of objects (gray dots) has a best-fit slope of 1.32$\pm$0.14 below 1.5 km. The black dots represent typical error bars on the total number of objects at each size range.}
\end{figure}

\clearpage

\begin{deluxetable}{lllllllllll}
\tabletypesize{\tiny}
\tablecolumns{11}
\tablecaption{NEATM results for the 428 NEOs detected by NEOWISE during the fully cryogenic portion of the WISE mission. An example table is given here; the full table is available electronically. This table contains the preliminary thermal fit results based on the First Pass version of the WISE data processing as described in the text. The NEOWISE project plans to produce an updated final catalog of physical properties based on the Second Pass processing of the WISE
 data using the updated version of the WISE Science Data System, with a goal of delivering this updated catalog to NASA's Planetary Data System.  Error values presented here represent the statistical errors on the model fits, including Monte Carlo modeling of uncertainties for the WISE magnitudes, $H$, $G$, and beaming and $p_{IR}$ when these two parameters cannot be fit.  Two calibration papers \citep{Mainzer11b,Mainzer11c} discuss the absolute calibration of the WISE data for small Solar system bodies and should be consulted before comparing with data derived from other sources.  The quoted precision for each parameter follows the object with the most significant figures for the error on that value in the  table. $H$, $G$ and albedo values of nan or ``-9.99" indicate that the objects have not received visible light follow-up.  Beaming value errors of nan or ``-9.99" indicate that the thermal fit routine returned a maximum value of pi or a minimum of 0.3, so error cannot be properly determined.  Readers are encouraged to check the WISE Explanatory Supplement \citep{Cutri} for details and updates.}
\tablehead{\colhead{Object} & \colhead{$H$} &\colhead{$G$} & \colhead{$D$ (km)} & \colhead{\pv}  & \colhead{$\eta$} & \colhead{$p_{IR}$} & \colhead{N(W1)} & \colhead{N(W2)} & \colhead{N(W3)} & \colhead{N(W4)} }
\startdata
K07X10C & 19.40 & 0.15 &    1.047 $\pm$    0.198 &    0.028 $\pm$    0.014 &    1.350 $\pm$    0.471  &    0.045 $\pm$    0.022 &   0 &   0 &  13 &   0\\
K10HA8Z & 21.00 & 0.15 &    0.356 $\pm$    0.067 &    0.055 $\pm$    0.032 &    1.350 $\pm$    0.475  &    0.089 $\pm$    0.051 &   0 &   0 &  11 &   0\\
  85713 & 15.70 & 0.15 &    3.484 $\pm$    0.789 &    0.076 $\pm$    0.039 &    1.350 $\pm$    0.450  &    0.122 $\pm$    0.062 &   0 &   0 &   9 &   0\\
  07822 & 17.40 & 0.15 &    1.208 $\pm$    0.015 &    0.133 $\pm$    0.022 &    2.261 $\pm$    0.049  &    0.299 $\pm$    0.219 &  23 &  24 &  24 &  24\\
  07822 & 17.40 & 0.15 &    1.602 $\pm$    0.012 &    0.075 $\pm$    0.014 &    2.171 $\pm$    0.030  &    0.139 $\pm$    0.043 &   9 &  10 &  11 &  11\\
K02N16W & 18.00 & 0.15 &    0.846 $\pm$    0.009 &    0.156 $\pm$    0.033 &    2.118 $\pm$    0.066  &    0.262 $\pm$    0.156 &  30 &  40 &  40 &  40\\
K09V24O & 19.80 & 0.15 &    0.467 $\pm$    0.016 &    0.098 $\pm$    0.020 &    1.964 $\pm$    0.130  &    0.157 $\pm$    0.167 &   0 &   5 &   5 &   5\\
K10MB2U & 20.60 & 0.15 &    0.599 $\pm$    0.022 &    0.028 $\pm$    0.006 &    1.684 $\pm$    0.111  &    0.045 $\pm$    0.134 &   0 &   7 &   7 &   7\\
  F4029 & 16.40 & 0.15 &    2.231 $\pm$    0.050 &    0.098 $\pm$    0.021 &    2.044 $\pm$    0.085  &    0.379 $\pm$    0.084 &   5 &   5 &   5 &   4\\
  F4029 & 16.40 & 0.15 &    2.164 $\pm$    0.228 &    0.104 $\pm$    0.028 &    1.396 $\pm$    0.261  &    0.166 $\pm$    0.045 &   0 &   0 &  12 &  10\\
  66251 & 17.00 & 0.15 &    1.222 $\pm$    0.233 &    0.187 $\pm$    0.094 &    1.350 $\pm$    0.419  &    0.300 $\pm$    0.150 &   0 &   0 &   4 &   0\\
\enddata
\end{deluxetable}
 

\clearpage
 \begin{thebibliography}{}
 
 \bibitem[Abell et al.(2009)]{Abell}
 Abell, P., Korsmeyer, D., Landis, R., Jones, T., Adamo, D., Morrison, D., Lemke, L., Gonzales, et al. 2009 Meteoritics \& Planetary Science 44, 1825
  
 \bibitem[Alvarez et al.(1980)]{Alvarez}
 Alvarez, L., Alvarez, W., Asaro, F., Michel, H. 1980 Science 208, 1095
 
 \bibitem[Bauer et al.(2011a)]{Bauer11a}
 Bauer, J. M., Walker, R. W., Mainzer, A. K., Masiero, J., R., Grav, T., et al. 2011 ApJ accepted
 
 \bibitem[Bauer et al.(2011b)]{Bauer11b}
 Bauer, J. M., Walker, R. W., Mainzer, A. K., Masiero, J., R., Grav, T., et al. 2011 in prep 
 
 \bibitem[Bhattacharya et al.(2010)]{Bhattacharya}
 Bhattacharya, B., Noriega-Crespo, A., Penprase, B., Meadows, V., Salvato, M., Aussel, H., Frayer, D., Ilbert, O., et al. 2010 \apj, 720, 113
 
\bibitem[Boslough \& Harris(2008)]{BosloughHarris}
 Boslough, M. B.; Harris, A. W. 2008 AGU U41D-0034
 
 \bibitem[Bottke et al.(2002)]{Bottke}
 Bottke, W., Morbidelli, A., Jedicke, R., Petit, J.-M., Levison, H., Michel, P., Metcalfe, T., 2002 Icarus, 156, 399
 
 \bibitem[Bowell et al.(1989)]{Bowell}
 Bowell, E., Hapke, B., Domingue, D., Lumme, K., Peltoniemi, J., Harris, A. Application of photometric models to asteroids.  In \emph{Asteroids II} (eds. R. P. Binzel, T. Gehrels and M. S. Matthews) pp. 524-556 University of Arizona Press, Tucson, Arizona
 
 \bibitem[Chapman(2004)]{Chapman}
 Chapman, C. 2004 Earth and Planetary Science Letters, 222, 1
 
\bibitem[Connors et al.(2011)]{Connors}
Connors, M., Wiegert, P., Veillet, C. 2011 Nature accepted

\bibitem[Cutri et al.(2011)]{Cutri}
 Cutri, R. M., Wright, Wright, E. L.; Conrow, T.; Bauer, J.; Benford, D.; Brandenburg, H.; Dailey, J.; Eisenhardt, P. R. M., et al. Explanatory Supplement to the WISE Preliminary Data Release Products, 2011 wise.rept
 
 \bibitem[Delb\'{o} et al.(2003)]{Delbo}
 Delb\'{o}, M., Harris, A. W., Binzel, R. P., Pravec, P., Davies, J. K. 2003 Icarus 166, 116
 
 \bibitem[Duncan \& Levison(1997)]{DuncanLevison}
 Duncan, M., \& Levison, H., 1997 Science 276, 1670
 
  \bibitem[Gaffey et al.(2002)]{Gaffey02}
 Gaffey, M., Cloutis, E. A., Kelley, M. S., Reed, K. L. in Asteroids III.  Bottke, W. F., Cellino, A., Paolicchi, P., Binzel, R. P., Eds. 2002 University of Arizona Press, Tuscon, AZ.

 \bibitem[Grav et al.(2011)]{Grav}
 Grav, T., Jedicke, T., Denneau, L, Chesley, S., Holman, M., Spahr, T. 2011 in press
 
  \bibitem[Harris(1998)]{Harris}
 Harris, A., 1998 Icarus, 131, 291
 
  \bibitem[Harris(2006)]{HarrisACM}
 Harris, A. W. Asteroids, Comets, Meteors, Proceedings of the 229th Symposium of the International Astronomical Union held in Bœzios, Rio de Janeiro, Brasil August 7-12, 2005, Edited by Daniela, L.; Sylvio Ferraz, M.; Angel, F. Julio Cambridge: Cambridge University Press, 2006., pp.449-463
 
 \bibitem[Harris(2008)]{Harris08}
 Harris, A., 2008 Nature 453, 1178
  
  \bibitem[Harris et al.(2009)]{Harris09}
 Harris, A., Mueller, M., Lisse, C., Cheng, A. 2009 Icarus, 199, 86
 
 \bibitem[Harris et al.(2011)]{Harris11}
 Harris, A., Mommert, M.; Hora, J. L.; Mueller, M.; Trilling, D. E.; Bhattacharya, B.; Bottke, W. F.; Chesley, S.; Delbo, M.; Emery, J. P.; et al. 2011 AJ 141, 75
 
 \bibitem[Helin et al.(1997)]{Helin}
Helin, E., Pravdo, S., Rabinowitz, D., Lawrence, K. 1997, Near-Earth Objects, the United Nations International Conference: Proceedings of the international conference held April 24-26, 1995 in New York, New York, USA. Edited by John L. Remo, 1997. Annals of the New York Academy of Sciences, vol. 822, p. 6

\bibitem[Hildebrand et al.(1991)]{Hildebrand}
Hildebrand, A.; Penfield, G.; Kring, D.; Pilkington, M.; Camargo Z., Antonio; J., Stein B.; Boynton, W. 1991 Geology 19, 867

\bibitem[Jedicke et al.(2002)]{Jedicke}
Jedicke, R., Larsen, J., Spahr, T. in Asteroids III.  Bottke, W. F., Cellino, A., Paolicchi, P., Binzel, R. P., eds. 2002 University of Arizona Press, Tucson, AZ. 

 \bibitem[Juri\'c et al.(2002)]{Juric}
 Juri\'c, M., et al. 2002, \aj, 124, 1776

 \bibitem[Kaasalainen et al.(2004)]{Kaasalainen}
 Kaasalainen, M., Pravec, P., Krugly, Y., et al. 2004 Icarus, 167, 178
 
 \bibitem[Kubica et al.(2007)]{Kubica}
 Kubica, J., Denneau, L., Grav, T., Heasley, J., Jedicke, R., et al. 2007 Icarus 189, 151
 
 \bibitem[Kasuga, Balam \& Wiegert(2010)]{Kasuga}
 Kasuga, T., Balam, D., Wiegert, P. 2010 AJ 140, 1806
 
 \bibitem[Koehn \& Bowell(2000)]{Koehn}
 Koehn, B. W.; Bowell, E. L. G. 2000. AAS/DPS Mtg 32, 14.03; BAAS 32, 1018.
 
 \bibitem[Larson(2007)]{Larson}
Larson, S., Near Earth Objects, our Celestial Neighbors: Opportunity and Risk, Proceedings if IAU Symposium 236. Edited by G.B. Valsecchi and D. Vokrouhlicky, and A. Milani. Cambridge: Cambridge University Press, 2007., pp.323-328
 
\bibitem[Lebofsky et al.(1978)]{Lebofsky78}
Lebofsky, L., Veeder, G., Lebofsky, M., Matson, D., 1978 Icarus, 35, 336

 \bibitem[Lebofsky \& Spencer(1989)]{Lebofsky_Spencer}
Lebofsky, L., \& Spencer, J., Asteroids II, University of Arizona Press, 1989, p. 128-147

 \bibitem[Levison \& Duncan(1997)]{LevisonDuncan}
 Levison, H., \& Duncan, M. 1997 Icarus 127, 13
 
 \bibitem[Levison et al.(1997)]{Levison97}
 Levison, H., Shoemaker, E., \& Shoemaker, C., 1997 Nature 385, 42
 
 \bibitem[Liu et al.(2008)]{Liu}
  Liu, F.; Cutri, R.; Greanias, G.; Duval, V.; Eisenhardt, P.; Elwell, J.; Heinrichsen, I.; Howard, J.; Irace, W.; Mainzer, A.; Razzaghi, A.; Royer, D.; Wright, E. L., 2008, SPIE, 7017, 16
 
 \bibitem[Luu \& Jewitt(1998)]{LuuJewitt}
 Luu, J., Jewitt, D., 1998 AJ 98, 1905
 
 \bibitem[Mainzer et al.(2005)]{Mainzer}
Mainzer, A.; Eisenhardt, P.; Wright, E. L.; Liu, F.; Irace, W.; Heinrichsen, I.; Cutri, R.; Duval, V.  2005, SPIE, 5899, 262

\bibitem[Mainzer et al.(2011a)]{Mainzer11a}
Mainzer, A., J. Bauer, T. Grav, J. Masiero, R. McMillan, R. Cutri, J. Dailey, E. Wright, et al., 2011a \apj 731, 53

\bibitem[Mainzer et al.(2011b)]{Mainzer11b}
Mainzer, A., Grav, T., Masiero, J., Bauer, J., et al. 2011b ApJ 736, 100

\bibitem[Mainzer et al.(2011c)]{Mainzer11c}
Mainzer, A., Grav, T., Masiero, J., Bauer, J., et al. 2011c ApJL 737, L9

\bibitem[Mainzer et al.(2011d)]{Mainzer11d}
Mainzer, A., Grav, T., Masiero, J., Hand, E., et al. 2011d ApJ accepted

\bibitem[Mainzer et al.(2011e)]{Mainzer11e}
Mainzer, A., Grav, T., Masiero, J., Bauer, J., et al. 2011e ApJ submitted

\bibitem[Masiero et al.(2011)]{Masiero}
Masiero, J., Mainzer, A., Grav, T., Bauer, J., et al. 2011 ApJ accepted

\bibitem[Matson(1986)]{Matson}
Matson, D., ed. The IRAS Asteroid and Comet Survey, 1986, JPL D-3698 (Pasadena: JPL).

\bibitem[McMillan(2007)]{McMillan}
McMillan, R. S., Near Earth Objects, our Celestial Neighbors: Opportunity and Risk, Proceedings if IAU Symposium 236. Edited by G.B. Valsecchi and D. Vokrouhlicky, and A. Milani. Cambridge: Cambridge University Press, 2007., pp.329-340
 
 \bibitem[Morbidelli \& Gladman(1998)]{MorbidelliGladman}
 Morbidelli, A., \& Gladman, B., 1998 M\&PS, 33, 999
 
 \bibitem[Morbidelli et al.(2002)]{Morbi02}
 Morbidelli, A., Jedicke, R., Bottke, W., Michel, P., Tedesco, E., 2002 Icarus 158, 329
 
 \bibitem[Morrison(1992)]{Morrison}
 Morrison, D. The Spaceguard Survey: Report of the NASA International Near Earth Object Detection Workshop. QB651 N37, Jet Propulsion Laboratory/California Institute of Technology, Pasadena.
 
 \bibitem[Mueller et al.(2011)]{Mueller}
 Mueller, M. Delbo, M., Hora, J., Trilling, D., Bhattacharya, B., Bottke W., Chesley, S., Emery, J., et al. 2011 AJ 141, 109
 
 \bibitem[National Research Council Report(2010)]{NRC}
 National Research Council Report, 2010 ``Defending Planet Earth"
 
 \bibitem[NEO Science Definition Team(2003)]{NEOSDT}
 Near-Earth Object Science Definition Team, 2003 ``Study to Determine the Feasibility of Extending the Search for Near-Earth Objects to Smaller Limiting Diameters"
 
 \bibitem[Parker et al.(2008)]{Parker}
Parker, A., Ivezi\'c, \v{Z}., Juri\'c, M., Lupton, R., Sekora, M., Kowalski, A. 2008 Icarus, 198, 138

\bibitem[Pravec et al.(2006)]{Pravec}
Pravec, P., Scheirich, P., Kusnirak, P., Sarounova, L., Mottola, S., Hahn, G., Brown, P., Esquerdo, G., et al. 2006 Icarus 181, 63
 
 \bibitem[Rabinowitz(1997a)]{Rabinowitz97a}
 Rabinowitz, D., 1997a Icarus 127, 33
 
 \bibitem[Rabinowitz(1997b)]{Rabinowitz97b}
 Rabinowitz, D., 1997b Icarus 130, 287
 
 \bibitem[Rabinowitz et al.(2000)]{Rabinowitz00}
 Rabinowitz, D., Helin, E., Lawrence, K., Pravdo, S. 2000 Nature 403, 165
 
\bibitem[Rivkin et al.(1997)]{Rivkin}
Rivkin, A., Howell, E., Clark, B., Lebofsky, L., Britt, D. 1997 LPI 28, 1183
 
 \bibitem[Ryan \& Ryan(2008)]{Ryan}
 Ryan, E., \& Ryan, W., 2008 Proceedings of the 2008 AMOS Technical Conference
 
 \bibitem[Shoemaker \& Helin(1978)]{ShoemakerHelin}
 Shoemaker, G., \& Helin, E., 1978, NASA CP-2053, pp. 245-256. 
 
 \bibitem[Shoemaker(1983)]{Shoemaker}
 Shoemaker, E. M., 1983 Ann. Rev. Earth Planet Sci., 11, 461
 
 \bibitem[Shoemaker et al.(1990)]{Shoemaker90}
 Shoemaker, E., Wolfe, R., Shoemaker, C. 1990 Geological Society of America Special Paper 247.
 
 \bibitem[Stokes et al.(2000)]{Stokes}
Stokes, G., Evans, J., Viggh, H, Shelly, F., Pearce, E., 2000, \icarus, 148, 21
 
 \bibitem[Stuart \& Binzel(2004)]{StuartBinzel}
 Stuart, J., \& Binzel, R. 2004 Icarus 170, 295
 
 \bibitem[Tedesco et al.(1988)]{Tedesco}
Tedesco, E., Matson, D., Veeder, G., Lebofsky, L., 1988, Comets to Cosmology: Proceedings of the Third IRAS Conference, Springer Berlin, Heidelberg, Vol. 297, p. 19-26

\bibitem[Tedesco et al.(2002)]{Tedesco02}
Tedesco, E., Noah, P., Noah, M., Price, S. 2002 \aj, 123, 1056

\bibitem[Tholen(2009)]{Tholen}
Tholen, D. J., Eds., Asteroid Absolute Magnitudes V12.0. EAR-A-5-DDR-ASTERMAG-V12.0. NASA Planetary Data System, 2009.

\bibitem[Thomas et al.(2011)]{Thomas}
Thomas, P., A'Hearn, M., Belton, M., Carcich, Lisse, C., Melosh, H., Schultz, P., Sunshine, J., et al. 2011 LPSC 1741

\bibitem[Thomas \& Binzel(2010)]{ThomasBinzel}
Thomas, C., \& Binzel, R., 2010 Icarus 205, 419
  
\bibitem[Trilling et al.(2010a)]{Trilling}
Trilling, D., et al.  2010, \aj, 140, 770

\bibitem[Trilling et al.(2010b)]{TrillingDPS}
Trilling, David E.; Mueller, M.; Hora, J. L.; Harris, A. W.; Bhattacharya, B.; Bottke, W. F.; Chesley, S.; Delbo, M.; Emery, J. P.; Fazio, G.; et al. 2010 DPS 42, 5706

\bibitem[Veeder et al.(1989)]{Veeder89}
Veeder, G., Hanner, M. S.; Matson, D. L.; Tedesco, E. F.; Lebofsky, L. A.; Tokunaga, A. T. 1989 AJ, 97, 1211

\bibitem[Vernazza et al.(2008)]{Vernazza}
Vernazza, P., Binzel, R., Thomas, C., DeMeo, F., Bus, S., Rivkin, A., Tokunaga, A. 2008 Nature 454, 858

\bibitem[Warner, Harris \& Pravec(2009)]{Warner}
Warner, B., Harris, A., Pravec, P. 2009 Icarus 202, 134

\bibitem[Weissman(1996)]{Weissman96}
Weissman, P., 1996. San Francisco: Astronomical Society of the Pacific, \emph{Completing the Inventory of the Solar System} ASP Conf. Ser. 107 pp. 265-288

 \bibitem[Wetherill(1988)]{Wetherill88}
 Wetherill, G. W., 1988 Icarus, 76, 1
 
 \bibitem[Williams, G.(2011)]{Williams}
Williams, G., personal communication, 2011

\bibitem[Wright et al.(2010)]{Wright}
Wright, E. L., Eisenhardt, Peter R. M., Mainzer, Amy K., Ressler, Michael E., Cutri, Roc M., et al. 2010   \aj, 140, 1868

\bibitem[Wolters et al.(2008)]{Wolters}
Wolters, S., Green, S., McBride, N., Davies, J. 2008 Icarus 193, 535


 \end{thebibliography}
\end{document}